\documentclass[reprint, aps, pra, floatfix]{revtex4-2}

\usepackage{amsfonts}
\usepackage{amsmath}
\usepackage{amssymb}
\usepackage{mathtools}
\usepackage{braket}
\usepackage{bm}
\usepackage{dsfont}

\usepackage{graphicx}
\usepackage{float}
\usepackage{import}

\usepackage{hyperref}
\usepackage{dcolumn}

\usepackage{color}
\usepackage[usenames, dvipsnames]{xcolor}

\definecolor{MyBlue}{RGB}{31, 119, 180}
\definecolor{MyRed}{RGB}{214, 39, 40}
\definecolor{MyGreen}{RGB}{44, 160, 44}
\definecolor{MyCerulean}{RGB}{23, 190, 207}

\DeclareUnicodeCharacter{2212}{$-$}

\hypersetup{
  colorlinks=true,
  linkcolor=MyBlue,
  linkbordercolor=MyBlue,
  citecolor=MyGreen,
  citebordercolor=MyGreen,
  urlcolor=MyCerulean,
  urlbordercolor=MyCerulean
}

\newcommand{\ddt}[1][]{%
  \frac{\mathrm{d} #1}{\mathrm{d}t}
}

\begin{document}

\preprint{APS/123-QED}


\title{Two-Photon Resonance Fluorescence in a Three-Level Ladder-Type Atom}

\author{Jacob Ngaha$^{1,2}$}
\email{j.ngaha@auckland.ac.nz}
\author{Scott Parkins$^{1,3}$}
\author{Howard J. Carmichael$^{1,3}$}
\affiliation{
  $^{1}$The Dodd-Walls Centre for Photonic and Quantum Technologies, New Zealand\\
  $^{2}$Department of Mathematics, The University of Auckland, Private Bag 92019, Auckland, New Zealand\\
  $^{3}$Department of Physics, The University of Auckland, Private Bag 92019, Auckland, New Zealand
}

\date{\today}

\begin{abstract}
  In this work, we consider a three-level ladder-type atom driven by a coherent field, inspired by the experimental work of Gasparinetti et al. [\href{https://doi.org/10.1103/PhysRevA.100.033802}{Phys. Rev. A 100, 033802 (2019)}]. When driven on two-photon resonance, the atom is excited into its highest energy state $\ket{f}$ by absorbing two photons simultaneously. The atom then de-excites via a cascaded decay $\ket{f} \rightarrow \ket{e} \rightarrow \ket{g}$. Here we present a theoretical study of the atomic fluorescence spectrum where, upon strong coherent driving, the spectrum exhibits seven distinct frequencies corresponding to transitions amongst the atomic dressed states. We characterize the quantum statistics of the emitted photons by investigating the second-order correlation functions of the emitted field. We do so by considering the total field emitted by the atom and focusing on each of the dressed-state components, taking in particular a secular-approximation and deriving straightforward, transparent analytic expressions for the second-order auto- and cross-correlations.
\end{abstract}

\maketitle

\section{Introduction}

It is well known for a two-level atom that, upon strong coherent driving, the fluorescence spectrum exhibits three distinct peaks. These peaks, known as the Mollow triplet \cite{Mollow_1969_PR_PowerSpectrumLight, Mollow_1975_PRA_PureStateAnalysis}, are due to transitions amongst the atomic dressed states, which are a result of the atom interacting with a strongly excited, quantized electromagnetic field mode \cite{CohenTannoudji_1977_JoPBAaMP_DressedAtomDescription, CohenTannoudji_1979_PTRSA_AtomsStrongLight}. The two-level atom is a simple, yet standard, example in the quantum optics vernacular. However, as the number of allowed atomic levels increases, so too does the complexity of the system. But in what way? In this work, we extend the two-level atom by introducing one more level, and driving, strongly and coherently, a two-photon transition.

There has been extensive work on three-level atoms in the past, and they remain a system of interest to this day \cite{Vogel_1992_PRA_ResonanceFluorescenceThree, Ficek_1995_PRA_FluorescenceIntensitySqueezing, Chang_2022_SR_CircuitQuantumElectrodynamics, Schoell_2020_PRL_CruxUsingCascaded}. In particular, the work of Whitley and Stroud \cite{Whitley_1976_PRA_DoubleOpticalResonance} laid some of the foundations for the theory of three-level systems, including taking a cavity QED approach by deriving steady-state behavior and absorption spectra using an operator equation of motion approach. This work was then later expanded on by Narducci et al. \cite{Narducci_1990_PRA_SpontaneousEmissionAbsorption} and Manka et al. \cite{Manka_1991_PRA_SpontaneousEmissionAbsorption} for three-level atoms in the $\Xi$, $\Lambda$, and V configurations. Reference~\cite{Manka_1991_PRA_SpontaneousEmissionAbsorption} is especially relevant due to its explanation of the emission spectra through the use of a dressed-state analysis in the high-intensity limit.

In this work, however, we focus on the model presented by Gasparinetti et al.: a three-level, anharmonic-ladder-type atom driven at two-photon resonance, as is relevant to the superconducting circuit QED system they implemented experimentally  \cite{Gasparinetti_2017_PRL_CorrelationsEntanglementMicrowave, Gasparinetti_2019_PRA_TwoPhotonResonance}. At two-photon resonance, the atom can only be excited to its highest energy level via the simultaneous absorption of two photons. Given the ladder-type configuration, the atom then de-excites back to the ground state via a cascaded decay through the intermediate state.

The earlier experimental work of Gasparinetti et al. \cite{Gasparinetti_2017_PRL_CorrelationsEntanglementMicrowave} focused on the weak driving limit, where the fluorescence spectrum exhibits only two frequency components. There they presented auto- and cross-correlations between photon emissions of these two frequency components demonstrating strong antibunching or ``superbunching'' depending on the order of detection. In their later work \cite{Gasparinetti_2019_PRA_TwoPhotonResonance}, the driving strength was increased to study the emergence of multiple spectral components, similar to the Mollow triplet. Antibunching and bunching was then demonstrated in the frequency-filtered auto-correlations of the single-photon resonance and two-photon resonance, respectively.

In this paper we present a much more in-depth theoretical study of the steady-state dynamics, incoherent fluorescence spectrum, and photon correlations of the field emitted by a three-level, ladder-type atom when driven by a resonant and coherent external field, focusing on the strong driving limit. Additionally, we further extend upon previous work on the dressed states of a three-level atom \cite{Koshino_2013_PRL_ObservationThreeState, Braunstein_2009_JPBAMOP_DressedStatesAnalysis, Narducci_1990_PRA_SpontaneousEmissionAbsorption, Manka_1991_PRA_SpontaneousEmissionAbsorption, Ngaha_Msc_Thesis_2019} by examining the photon correlations of the frequency components of the fluorescence spectrum and taking a secular approximation. This allows us to derive analytic expressions for second-order correlations of the many dressed-state transitions, i.e., the different spectral components of the fluorescence spectrum.

We begin by introducing the Hamiltonian and master equation for the radiatively damped three-level ladder-type atom in Sec.~\ref{sect:2_dynamics}. We then present a numerical survey of the atomic steady states, comparing them to analytic expressions derived for a two-level approximation of the atom. In Sec.~\ref{sect:3_spectrum} we present a survey of the incoherent atomic power spectrum when driven at, or near, two-photon resonance. Finally, Sec.~\ref{sect:4_atomic_g2} is dedicated to the atomic photon correlations of the emitted field under weak and strong coherent driving. Taking a secular approximation \cite{Schrama_1992_PRA_IntensityCorrelationsComponents}, we then derive analytic expressions for the frequency-resolved photon correlations of each spectral resonance in the strong driving limit.

The code used to generate the data presented in this paper is openly available \cite{github_for_paper_data}, and makes use of the quantum toolbox in Python (QuTiP) \cite{Johansson_2012_CPC_QutipOpenSource, Johansson_2013_CPC_Qutip2Python, Lambert_2024_arXiv_Qutip5Quantum}.

\section{Atomic Dynamics of the Ladder-Type Atom}
\label{sect:2_dynamics}

\subsection{The atomic moment equations}

\begin{figure}
  \centering
  {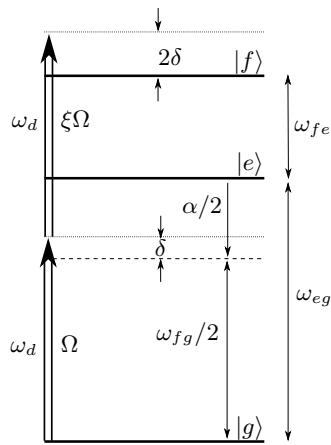}
  \caption{Energy level diagram of the three-level ladder-type atom: $\delta$ is the detuning of the drive frequency $\omega_{d}$ from the two-photon resonance frequency, $\omega_{fg} / 2$; $\alpha$ is the detuning of the driving field from the two single-photon resonance frequencies, $\omega_{eg}$ and $\omega_{fe}$; and $\Omega$ and $\xi \Omega$ are Rabi frequencies for the lower and upper dipole transitions, respectively.}
  \label{fig:2_atom_diagram}
\end{figure}

We consider a three-level ladder-type atom with ground state $\ket{g}$, intermediate state $\ket{e}$, and excited state $\ket{f}$, with respective energies $\hbar \omega_{g}, \hbar \omega_{e}$, and $\hbar \omega_{f}$. With coherent driving of the $\ket{g} \leftrightarrow \ket{e}$ and $\ket{e} \leftrightarrow \ket{f}$ transitions at driving frequency $\omega_{d}$, the Hamiltonian for the driven system is given by:
\begin{align}\label{eq:2_H_atom_time_indepdent}
  H_{A} &= \hbar \omega_{g} \ket{g} \bra{g} + \hbar \omega_{e} \ket{e} \bra{e} + \hbar \omega_{f} \ket{f} \bra{f} \nonumber \\
  &\quad + \hbar \frac{ \Omega }{ 2 } \left( e^{i \omega_{d} t} \ket{g} \bra{e} + e^{-i \omega_{d} t} \ket{e} \bra{g} \right) \nonumber \\
  &\quad + \hbar \xi \frac{ \Omega }{ 2 } \left( e^{i \omega_{d} t} \ket{e} \bra{f} + e^{-i \omega_{d} t} \ket{f} \bra{e} \right) ,
\end{align}
where $\xi$ is the ratio of the dipole moments for the upper and lower dipole transitions, and $\Omega$ is the coherent driving amplitude of the lower transition. We transform Eq.~(\ref{eq:2_H_atom_time_indepdent}) into a frame rotating at the drive frequency to obtain the time-independent Hamiltonian:
\begin{align}\label{eq:2_H_atom}
  H_{A} &= -\hbar \left( \frac{ \alpha }{ 2 } + \delta \right) \ket{e} \bra{e} - 2 \hbar \delta \ket{f} \bra{f} \nonumber \\
  &\quad\quad + \hbar \frac{ \Omega }{ 2 } \left( \Sigma_{+} + \Sigma_{-} \right),
\end{align}
where, following the notation of Gasparinetti et al. \cite{Gasparinetti_2017_PRL_CorrelationsEntanglementMicrowave, Gasparinetti_2019_PRA_TwoPhotonResonance},
\begin{equation}
  \alpha \equiv \omega_{fe} - \omega_{eg}
\end{equation}
is the \textit{anharmonicity} of the $\ket{e} \rightarrow \ket{g}$ and $\ket{f} \rightarrow \ket{e}$ transitions, with $\omega_{ij} = \omega_{i} - \omega_{j}$,
\begin{equation}
  \delta \equiv \omega_{d} - \frac{ \omega_{fg} }{ 2 }
\end{equation}
is the drive detuning from two-photon resonance, and
\begin{equation}\label{eq:2_atomic_raising_lowering_operators}
  \Sigma_{-} \equiv \ket{g} \bra{e} + \xi \ket{e} \bra{f}, \quad \Sigma_{+} \equiv \ket{e} \bra{g} + \xi \ket{f} \bra{e}
\end{equation}
are the atomic lowering and raising operators, respectively. Figure~\ref{fig:2_atom_diagram} depicts the structure and excitation of the three-level atom.

We model energy dissipation through spontaneous emission from the atom with the Lindblad master equation
\begin{equation}\label{eq:2_atomic_master_equation}
  \ddt[\rho] = \frac{ 1 }{ i \hbar } [ H_{A}, \rho ] + \frac{ \Gamma }{ 2 } \Lambda(\Sigma_{-}) \rho,
\end{equation}
where $\rho = \sum_{i,j = g, e, f} \rho_{ij} \ket{i} \bra{j}$ is the atomic density operator, $\Gamma$ is the population decay rate for the lower dipole, and $\Lambda(X)\bullet = \left( 2 X \bullet X^{\dagger} - X^{\dagger} X \bullet - \bullet X^{\dagger} X \right)$ is the Lindblad decay superoperator.

Considering the case of two-photon resonance -- where the atom is excited to the $\ket{f}$ state via a \emph{simultaneous} absorption of two photons -- we assume a large detuning between the two dipole transitions, i.e., a large anharmonicity $|\alpha|$ relative to the natural linewidth $\Gamma$; for $|\alpha| \sim \Gamma$, the excited state $\ket{f}$ can be populated through two separate absorption events via the intermediate state $\ket{e}$. The transmon qubit in the work of Gasparinetti et al. had an anharmonicity $\alpha / 2 \pi = -233$ MHz and a linewidth of $\Gamma / 2 \pi \approx 1.9$ MHz \cite{Gasparinetti_2019_PRA_TwoPhotonResonance, Gasparinetti_2019_PRA_TwoPhotonResonance}, therefore, in this work, we will set $\alpha / \Gamma = -120$. Changing the anharmonicity alters the overall structure of the atomic ladder structure; however, the key results remain largely unchanged.

\subsection{Survey of the atomic steady-state populations}

From the master equation Eq.~(\ref{eq:2_atomic_master_equation}), we can derive equations of motion for expectation values of a set of nine atomic operators. These can readily be expressed as a set of optical Bloch-like equations \cite{Narducci_1990_PRA_SpontaneousEmissionAbsorption, Whitley_1976_PRA_DoubleOpticalResonance}:
\begin{equation}\label{eq:2_bloch_like_equation}
  \ddt \langle \bm{\Sigma} \rangle = \bm{M} \langle \bm{\Sigma} \rangle + \bm{B},
\end{equation}
with $( \sigma_{ij} = \ket{i} \bra{j} )$:
\begin{equation}
  \langle \bm{\Sigma} \rangle =
  \begin{pmatrix}
    \langle \sigma_{gg} \rangle \\
    \langle \sigma_{ge} \rangle \\
    \langle \sigma_{eg} \rangle \\
    \langle \sigma_{ee} \rangle \\
    \langle \sigma_{ef} \rangle \\
    \langle \sigma_{fe} \rangle \\
    \langle \sigma_{gf} \rangle \\
    \langle \sigma_{fg} \rangle
  \end{pmatrix}
  ,\quad \bm{B} =
  \begin{pmatrix}
    0 \\
    0 \\
    0 \\
    \Gamma \xi^{2} \\
    i \xi \frac{ \Omega }{ 2 } \\
    -i \xi \frac{ \Omega }{ 2 } \\
    0 \\
    0
  \end{pmatrix} ,
\end{equation}
and
\begin{widetext}
  \begin{align}
    \bm{M}^{(\Sigma)} =
      \begin{psmallmatrix}
        0 & -i \frac{ \Omega }{ 2 } & i \frac{ \Omega }{ 2 } & \Gamma & 0 & 0 & 0 & 0 \\
        -i \frac{ \Omega }{ 2 } & -\left[ \frac{ \Gamma }{ 2 } - i \left( \frac{ \alpha }{ 2 } + \delta \right) \right] & 0 & i \frac{ \Omega }{ 2 } & \Gamma \xi & 0 & -i \xi \frac{ \Omega }{ 2 } & 0 \\
        i \frac{ \Omega }{ 2 } & 0 & -\left[ \frac{ \Gamma }{ 2 } + i \left( \frac{ \alpha }{ 2 } + \delta \right) \right] & -i \frac{ \Omega }{ 2 } & 0 & \Gamma \xi & 0 & i \xi \frac{ \Omega }{ 2 }\\
        -\Gamma \xi^{2} & i \frac{ \Omega }{ 2 } & -i \frac{ \Omega }{ 2 } & -\Gamma \left( 1 + \xi^{2} \right)& -i \xi \frac{ \Omega }{ 2 } & i \xi \frac{ \Omega }{ 2 } & 0 & 0 \\
        -i \xi \frac{ \Omega }{ 2 } & 0 & 0 & -i \xi \Omega & -\left[ \frac{ \Gamma }{ 2 } \left( 1 + \xi^{2} \right) + i \left( \frac{ \alpha }{ 2 } - \delta \right) \right] & 0 & i \frac{ \Omega }{ 2 } & 0 \\
        i \xi \frac{ \Omega }{ 2 } & 0 & 0 & i \xi \Omega & 0 & -\left[ \frac{ \Gamma }{ 2 } \left( 1 + \xi^{2} \right) - i \left( \frac{ \alpha }{ 2 } - \delta \right) \right] & 0 & -i \frac{ \Omega }{ 2 } \\
        0 & -i \xi \frac{ \Omega }{ 2 } & 0 & 0 & i \frac{ \Omega }{ 2 } & 0 & -\left[ \frac{ \Gamma }{ 2 } \xi^{2} - 2 i \delta \right] & 0 \\
        0 & 0 & i \xi \frac{ \Omega }{ 2 } & 0 & 0 & -i \frac{ \Omega }{ 2 } & 0 & -\left[ \frac{ \Gamma }{ 2 } \xi^{2} + 2 i \delta \right]
      \end{psmallmatrix} .
  \end{align}
\end{widetext}
Here we have used the fact that $\langle \sigma_{gg} \rangle + \langle \sigma_{ee}\rangle + \langle \sigma_{ff} \rangle = 1$ to reduce the set of nine coupled equations to eight. We solve for the steady-state expectation values of the atomic operators by numerically evaluating the steady-state vector of Eq.~(\ref{eq:2_bloch_like_equation}): $\langle \bm{\Sigma} \rangle_{ss} = -\bm{M}^{-1} \bm{B}$.

\begin{figure}
  \centering
  {\fontsize{7pt}{1em}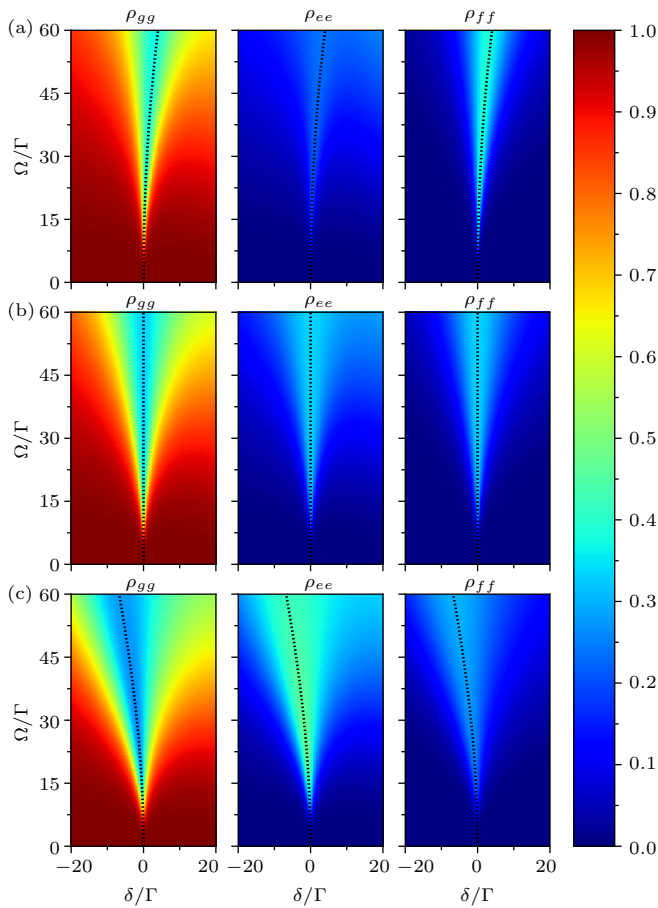}
  \caption{Steady-state populations from Eq.~(\ref{eq:2_atomic_master_equation}) of the $\ket{g}$ (left), $\ket{e}$ (center), and $\ket{f}$ (right) atomic states as a function of the drive detuning $\delta$ and drive amplitude $\Omega$ for three values of the dipole moment ratio: (a) $\xi = 1 / \sqrt{2}$, (b) $\xi = 1$, and (c) $\xi = \sqrt{2}$. Also shown is the Stark-shifted two-photon resonance frequency (black, dotted), Eq.~(\ref{eq:2_shifted_resonance}). The atomic anharmonicity is $\alpha = -120 \Gamma$.}
  \label{fig:2_steady_states_map}
\end{figure}

In Fig.~\ref{fig:2_steady_states_map}, we depict the steady-state populations of the ground, intermediate, and excited states as a function of the drive amplitude and detuning for three values of the dipole moment ratio: $\xi = 1 / \sqrt{2}$ [Fig.~\ref{fig:2_steady_states_map}(a)], $\xi = 1$ [Fig.~\ref{fig:2_steady_states_map}(b)], and $\xi = \sqrt{2}$ [Fig.~\ref{fig:2_steady_states_map}(c)].

As one might expect, the atom remains in the ground state for weak driving amplitudes far from two-photon resonance. We only see significant population in the $\ket{e}$ and $\ket{f}$ states for strong, near-resonant driving, when the atom can simultaneously absorb two photons. For all three values of $\xi$, there is a slight shift in population towards positive detuning, i.e., towards the single-photon resonance at $\delta = -\alpha / 2$.

It is clear that the dipole moment ratio also plays a significant role in the evolution of the atom; it determines the ratio of the driving strengths and the respective decay rates between the two dipole moments. When $\xi = 1$, both of these ratios are the same, and we see qualitatively similar steady-state populations in the $\ket{e}$ and $\ket{f}$ states in Fig.~\ref{fig:2_steady_states_map}(b). When $\xi < 1$, there is a significantly higher population in the $\ket{f}$ state than in the $\ket{e}$ state. Although the upper dipole moment is weaker than the lower dipole moment, the ratio of the driving amplitudes scales with $\xi$, while the ratio of the decay rates scales with $\xi^{2}$, leading to a higher population in the excited state $\ket{f}$. We see the opposite effect when $\xi > 1$ in Fig.~\ref{fig:2_steady_states_map}(c), with a faster decay rate for the upper dipole moment, resulting in a larger steady-state population in the $\ket{e}$ state.

For $\xi = 1$, the steady-state atomic populations are roughly symmetric about two-photon resonance, with a slight shift towards the single-photon resonance as previously mentioned. However, this symmetry is broken, when $\xi \neq 1$, with an apparent resonance shift towards positive detunings for $\xi < 1$ and towards negative detunings for $\xi > 1$. As the drive strength increases, there is an induced level shift of the $\ket{g}, \ket{e}$, and $\ket{f}$ states, causing a shift in two-photon resonance frequency. This shifted resonance -- Eq.~(\ref{eq:2_shifted_resonance}) in the following section -- is shown on top of each of the population maps in Fig.~\ref{fig:2_steady_states_map}. Equation~(\ref{eq:2_shifted_resonance}) clearly indicates that, when $\xi = 1$, there is no overall shift in the two-photon resonance as the driving strength $\Omega$ is varied.

\subsection{Effective two-level model}

Given the relatively low population in the intermediate state, we take a secular approximation and adiabatically eliminate the $\ket{e}$ state from the Hamiltonian to obtain an effective ``two-level model'' \cite{Sinatra_1995_QaSOJotEOSPB_EffectiveTwoLevel, Wu_1997_PRA_EffectiveTwoLevel}. We assume the atom has a large anharmonicity with a driving frequency close to two-photon resonance, such that $|\alpha| \gg \delta$. The Hamiltonian for this effective two-level model is then given by:
\begin{align}
  H_{\mathrm{eff}} &= \hbar \Delta_{g} \ket{g} \bra{g} + \hbar \left( -2 \delta + \Delta_{f} \right) \ket{f} \bra{f} \nonumber \\
  &\quad + \hbar \frac{ \Omega_{\mathrm{eff}} }{ 2 } \Big( \ket{g} \bra{f} + \ket{f} \bra{g} \Big) .
\end{align}
where
\begin{equation}\label{eq:2_effective_two_photon_drive}
  \frac{ \Omega_{\mathrm{eff}} }{ 2 } \equiv \xi \left( \frac{ \Omega }{ 2 } \right)^{2} \frac{ 1 }{ \alpha / 2  + \delta }
\end{equation}
is the \textit{effective} two-photon driving amplitude, and
\begin{subequations}
  \begin{align}
    \Delta_{g} &\equiv \left( \frac{ \Omega }{ 2 } \right)^{2} \frac{ 1 }{ \alpha / 2 + \delta }, \\
    \Delta_{f} &\equiv \left( \frac{ \xi \Omega }{ 2 } \right)^{2} \frac{ 1 }{ \alpha / 2 + \delta } ,
  \end{align}
\end{subequations}
are the induced Stark shifts to the ground and excited states, respectively (see Appendix \ref{app:1_effective_2LA_model} for more details). These Stark shifts arise from virtual transitions to the intermediate state $\ket{e}$ via simultaneous emission and absorption at the single-photon resonances, $\omega_{eg}$ and $\omega_{fe}$. The Stark shifts cause a corresponding shift in the two-photon resonance frequency; therefore, we define the \textit{effective} drive detuning from two-photon resonance as
\begin{equation}\label{eq:2_effective_detuning}
  \Delta_{\mathrm{eff}} \equiv \left( -2 \delta + \Delta_{f} \right) - \Delta_{g} = -2 \delta + \left( \frac{ \Omega }{ 2 } \right)^{2} \frac{ \xi^{2} - 1 }{ \alpha / 2 + \delta }.
\end{equation}
By rearranging Eq.~(\ref{eq:2_effective_detuning}) at resonance, $\Delta_{\mathrm{eff}} = 0$, we find the shifted two-photon resonance detuning is
\begin{equation}\label{eq:2_shifted_resonance}
  \delta_{\mathrm{shifted}} \equiv -\frac{ \alpha }{ 4 } - \frac{ 1 }{ 2 } \sqrt{ \left( \frac{ \alpha }{ 2 } \right)^{2} + 2 \left( \frac{ \Omega }{ 2 } \right)^{2} \left( \xi^{2} - 1 \right) } ,
\end{equation}
which, as previously mentioned, is shown in Fig.~\ref{fig:2_steady_states_map}.

\begin{figure*}
  \centering
  {\fontsize{7pt}{0pt}{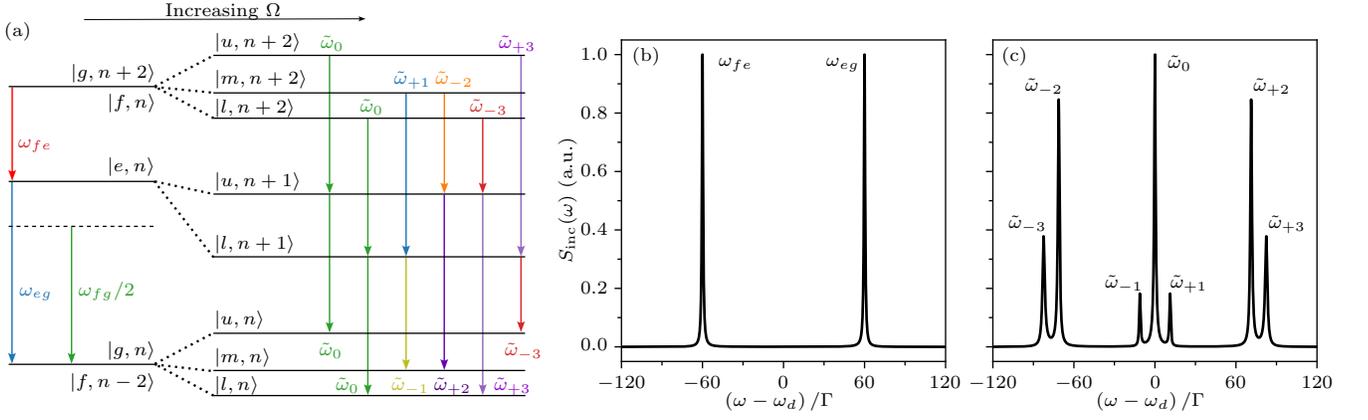}}
  \caption{(a) Energy level diagram of the dressed states of the three-level ladder-type atom at two-photon resonance ($\delta = 0$), with all possible de-excitation paths shown. As the driving amplitude increases, the degeneracy of the dressed states is lifted, allowing for a total of twelve de-excitation paths. (b) Incoherent power spectrum in the weak driving, $\Omega = 5 \Gamma$, and (c) strong driving, $\Omega = 40 \Gamma$, regimes. Each peak is labeled with the appropriate transition frequencies, Eq.~(\ref{eq:3_dressed_state_frequencies}). The other parameters are $\alpha = -120 \Gamma, \delta = 0$, and $\xi = 1$.}
  \label{fig:3_dressed_states_labelled_spectrum}
\end{figure*}

Assuming a large anharmonicity, we can drop the rapidly oscillating cross terms in the Lindblad decay term of the master equation, Eq.~(\ref{eq:2_atomic_master_equation}), resulting in a master equation with two separate decay terms corresponding to decays from each dipole moment $\left( \sigma_{ij} = \ket{i} \bra{j} \right)$:
\begin{equation}\label{eq:2_effective_model_master_equation}
  \ddt[\rho] = \frac{ 1 }{ i \hbar } [H_{\mathrm{eff}}, \rho ] + \frac{ \Gamma }{ 2 } \Lambda(\sigma_{ge}) \rho + \frac{ \xi^{2} \Gamma }{ 2 } \Lambda(\sigma_{ef}) \rho .
\end{equation}
Equation~(\ref{eq:2_effective_model_master_equation}) can then be rewritten as two sets of coupled ordinary differential equations for the components of the reduced density operator \cite{Whitley_1976_PRA_DoubleOpticalResonance}:
\begin{subequations}
  \begin{align}
    \ddt \langle \sigma_{gg} \rangle &= -\Gamma \langle \sigma_{gg} \rangle - \Gamma \langle \sigma_{ff} \rangle - i \frac{ \Omega_{\mathrm{eff}} }{ 2 } \langle \sigma_{gf} \rangle \nonumber \\
    &\quad + i \frac{ \Omega_{\mathrm{eff}} }{ 2 } \langle \sigma_{fg} \rangle + \Gamma , \\
    \ddt \langle \sigma_{ff} \rangle &= -\xi^{2} \Gamma \langle \sigma_{ff} \rangle + i \frac{ \Omega_{\mathrm{eff}} }{ 2 } \langle \sigma_{gf} \rangle - i \frac{ \Omega_{\mathrm{eff}} }{ 2 } \langle \sigma_{fg} \rangle, \\
    \ddt \langle \sigma_{gf} \rangle &= -i \frac{ \Omega_{\mathrm{eff}} }{ 2 } \langle \sigma_{gg} \rangle + i \frac{ \Omega_{\mathrm{eff}} }{ 2 } \langle \sigma_{ff} \rangle \nonumber \\
    &\quad - \left( \frac{ \xi^{2} \Gamma }{ 2 } + i \Delta_{\mathrm{eff}} \right) \langle \sigma_{gf} \rangle , \\
    \ddt \langle \sigma_{fg} \rangle &= i \frac{ \Omega_{\mathrm{eff}} }{ 2 } \langle \sigma_{gg} \rangle - i \frac{ \Omega_{\mathrm{eff}} }{ 2 } \langle \sigma_{ff} \rangle \nonumber \\
    &\quad - \left( \frac{ \xi^{2} \Gamma }{ 2 } - i \Delta_{\mathrm{eff}} \right) \langle \sigma_{fg} \rangle ,
  \end{align}
\end{subequations}
and
\begin{subequations}
  \begin{align}
    \ddt \langle \sigma_{ge} \rangle &= -\left( \frac{ \Gamma }{ 2 } - i \Delta_{g} \right) \langle \sigma_{ge} \rangle + i \frac{ \Omega_{\mathrm{eff}} }{ 2 } \langle \sigma_{fe} \rangle , \\
    \ddt \langle \sigma_{eg} \rangle &= -\left( \frac{ \Gamma }{ 2 } + i \Delta_{g} \right) \langle \sigma_{eg} \rangle - i \frac{ \Omega_{\mathrm{eff}} }{ 2 } \langle \sigma_{ef} \rangle , \\
    \ddt \langle \sigma_{ef} \rangle &= -\left[ \frac{ \Gamma }{ 2 } \left( 1 + \xi^{2} \right) - i \left( 2\delta - \Delta_{f} \right) \right] \langle \sigma_{ef} \rangle \nonumber \\
    &\quad - i \frac{ \Omega_{\mathrm{eff}} }{ 2 } \langle \sigma_{eg} \rangle, \\
    \ddt \langle \sigma_{fe} \rangle &= -\left[ \frac{ \Gamma }{ 2 } \left( 1 + \xi^{2} \right) + i \left( 2\delta - \Delta_{f} \right) \right] \langle \sigma_{ef} \rangle \nonumber \\
    &\quad + i \frac{ \Omega_{\mathrm{eff}} }{ 2 } \langle \sigma_{ge} \rangle .
  \end{align}
\end{subequations}
Solving for the steady-state atomic state populations, we find:
\begin{subequations}\label{eq:2_approximate_steady_states}
  \begin{align}
    \langle \sigma_{gg} \rangle_{ss} &= \frac{ \Omega_{\mathrm{eff}}^{2} + 4 \Delta_{\mathrm{eff}}^{2} + \xi^{4} \Gamma^{2} }{ \Omega_{\mathrm{eff}}^{2} \left( 2 + \xi^{2} \right) + 4 \Delta_{\mathrm{eff}}^{2} + \xi^{4} \Gamma^{2} } , \\
    \langle \sigma_{ee} \rangle_{ss} &= \frac{ \xi^{2} \Omega_{\mathrm{eff}}^{2} }{ \Omega_{\mathrm{eff}}^{2} \left( 2 + \xi^{2} \right) + 4 \Delta_{\mathrm{eff}}^{2} + \xi^{4} \Gamma^{2} } , \\
    \langle \sigma_{ff} \rangle_{ss} &= \frac{ \Omega_{\mathrm{eff}}^{2} }{ \Omega_{\mathrm{eff}}^{2} \left( 2 + \xi^{2} \right) + 4 \Delta_{\mathrm{eff}}^{2} + \xi^{4} \Gamma^{2} } .
  \end{align}
\end{subequations}
Assuming a strong effective two-photon drive $\left( \Omega_{\mathrm{eff}} \gg \Gamma \right)$, a large anharmonicity $\left( |\alpha| \gg \Gamma \right)$, and a drive frequency close to two-photon resonance $\left( |\alpha|/2 \gg |\delta| \right)$, these analytic expressions give close agreement to the full model as computed from Eq.~(\ref{eq:2_bloch_like_equation}). In particular, we see the effect of the dipole moment ratio $\xi$ on the steady-state population of the excited state $\ket{f}$ shown in Fig.~\ref{fig:2_steady_states_map}. From Eq.~(\ref{eq:2_approximate_steady_states}), the ratio of the $\ket{e}$ and $\ket{f}$ steady state populations is $\langle \sigma_{ee} \rangle_{ss} / \langle \sigma_{ff} \rangle_{ss}  = \xi^{2},$
hence, for $\xi > 1$, there is more population in the $\ket{e}$ state than in the $\ket{f}$ state, and vice versa for $\xi < 1$.

\section{Atomic Fluorescence and the Optical Spectrum}
\label{sect:3_spectrum}

\subsection{The atomic dressed states}

By diagonalizing the Hamiltonian, Eq.~(\ref{eq:2_H_atom_time_indepdent}), we find that the dressed-state eigenfrequencies are solutions to the characteristic polynomial,
\begin{align}\label{eq:3_characteristic_polynomial}
  \omega_{i}^{3} + &\left( \frac{ \alpha }{ 2 } + 3 \delta \right) \omega_{i}^{2} + \nonumber \\
  &\left[ \delta \left( \alpha + 2 \delta \right) - \left( \frac{ \Omega }{ 2 } \right)^{2} \left( 1 + \xi^{2} \right) \right] \omega_{i} = 2 \delta \left( \frac{ \Omega }{ 2 } \right)^{2}.
\end{align}
We can simplify this cubic equation by considering the special case of two-photon resonance, setting $\delta = 0$. The eigenfrequencies are then \cite{Narducci_1990_PRA_SpontaneousEmissionAbsorption}
\begin{subequations}\label{eq:3_dressed_state_frequencies}
  \begin{gather}
    \omega_{m} = 0 , \\
    \omega_{u} = -\frac{ \alpha }{ 4 } - \sqrt{ \left( \frac{ \alpha }{ 4 } \right)^{2} + \left( \frac{ \Omega }{ 2 } \right)^{2} \left( 1 + \xi^{2} \right) } , \\
    \omega_{l} = -\frac{ \alpha }{ 4 } + \sqrt{ \left( \frac{ \alpha }{ 4 } \right)^{2} + \left( \frac{ \Omega }{ 2 } \right)^{2} \left( 1 + \xi^{2} \right) } ,
  \end{gather}
\end{subequations}
with corresponding eigenvectors:
\begin{subequations}\label{eq:3_dressed_states}
  \begin{gather}
    \ket{m} = \frac{ -1 }{ \sqrt{ 1 + \xi^{2} } } \Big( \xi \ket{g} - \ket{f} \Big) , \\
    \ket{u} = \frac{ 1 }{ \sqrt{ 4 \omega_{u}^{2} + \Omega^{2} \left( 1 + \xi^{2} \right) } } \Big( \Omega \ket{g} + 2 \omega_{u} \ket{e} + \xi \Omega \ket{f} \Big) , \\
    \ket{l} = \frac{ 1 }{ \sqrt{ 4 \omega_{l}^{2} + \Omega^{2} \left( 1 + \xi^{2} \right) } } \Big( \Omega \ket{g} + 2 \omega_{l} \ket{e} + \xi \Omega \ket{f} \Big)  .
  \end{gather}
\end{subequations}
These three dressed states, depicted in Fig.~\ref{fig:3_dressed_states_labelled_spectrum}(a), along with the uncoupled atomic states, open up several possible de-excitation channels. There are, in fact, twelve possible transitions in total, with seven distinct frequencies appearing:
\begin{subequations}\label{eq:3_transition_frequencies}
  \begin{align}
    \tilde{\omega}_{0} &= 0, \\
    \tilde{\omega}_{\pm 1} &= \pm \left( \omega_{m} - \omega_{l} \right) , \\
    \tilde{\omega}_{\pm 2} &= \pm \left( \omega_{u} - \omega_{m} \right) , \\
    \tilde{\omega}_{\pm 3} &= \pm \left( \omega_{u} - \omega_{l} \right) .
  \end{align}
\end{subequations}
Figure~\ref{fig:3_dressed_states_labelled_spectrum}(a) also shows that the possible orderings of the allowed dressed-state transitions follow a cascaded decay, as in the bare state picture.

\subsection{Incoherent power spectrum near two-photon resonance}

We define the normalized first-order correlation function for the three-level ladder-type atom in the steady-state limit as:
\begin{align}
  g^{(1)}_{ss}(\tau) &= \lim_{t \rightarrow \infty} \frac{ \langle \Sigma_{+}(t + \tau) \Sigma_{-}(t) \rangle }{ \sqrt{ \langle \Sigma_{+} \Sigma_{-}(t) \rangle \langle \Sigma_{+} \Sigma_{-}(t + \tau) \rangle } } \nonumber \\
  &= \frac{ \langle \Sigma_{+}(\tau) \Sigma_{-}(0) \rangle }{ \langle \Sigma_{+} \Sigma_{-} \rangle_{ss} } .
\end{align}
From the Wiener-Khinchin theorem \cite{Khinchin_1934_MA_KorrelationstheorieDerStationaren, Wiener_1930_AM_GeneralizedHarmonicAnalysis}, we can decompose the power spectrum into coherent and incoherent components \cite{Whitley_1976_PRA_DoubleOpticalResonance}:
\begin{equation}
  S(\omega) = S_{\mathrm{coh}}(\omega) + S_{\mathrm{inc}}(\omega) ,
\end{equation}
with
\begin{subequations}
  \begin{align}
    S_{\mathrm{coh}}(\omega) &= \frac{ 1 }{ 2 \pi } \int_{-\infty}^{\infty} e^{i \omega \tau} \frac{ \langle \Sigma_{+} \rangle_{ss} \langle \Sigma_{-} \rangle_{ss} }{ \langle \Sigma_{+} \Sigma_{-} \rangle_{ss} } \mathrm{d}\tau , \\
    S_{\mathrm{inc}}(\omega) &= \frac{ 1 }{ 2 \pi } \int_{-\infty}^{\infty} e^{i \omega \tau} \frac{ \langle \Delta \Sigma_{+}(\tau) \Delta \Sigma_{-}(0) \rangle }{ \langle \Sigma_{+} \Sigma_{-} \rangle_{ss} } \mathrm{d}\tau ,
  \end{align}
\end{subequations}
where $\Delta \Sigma_{\pm} = \Sigma_{\pm} - \langle \Sigma_{\pm} \rangle_{ss}$ are the atomic raising and lowering quantum fluctuation operators.

\begin{figure}
  \centering
  {\fontsize{7pt}{1em}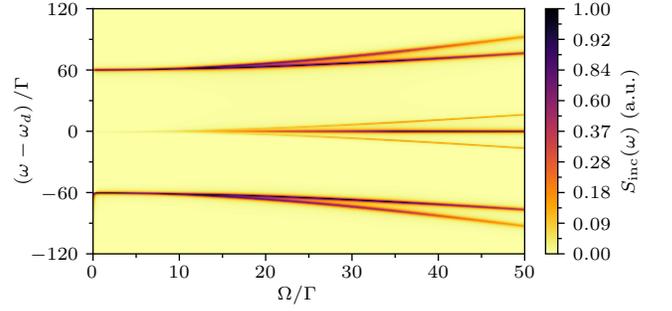}
  \caption{Normalized incoherent power spectrum as a function of driving amplitude $\Omega$ at two-photon resonance $\left( \delta = 0 \right)$. The other parameters are $\alpha = -120 \Gamma$ and $\xi = 1$.}
  \label{fig:3_spectrum_scan_Omega}
\end{figure}

A weakly driven two-level atom displays a single peak in its fluorescence spectrum as it decays from the excited state to the ground state. For the three-level atom, we observe two separate peaks corresponding to the cascaded decay from the two single-photon resonances: $\ket{f} \rightarrow \ket{e}$ followed by $\ket{e} \rightarrow \ket{g}$, as shown in Fig.~\ref{fig:3_dressed_states_labelled_spectrum}(b). A weak, off-resonant drive results in a weak effective two-photon drive, leading to a low population in the excited state $\ket{f}$.

\begin{figure}
  \centering
  {\fontsize{7pt}{1em}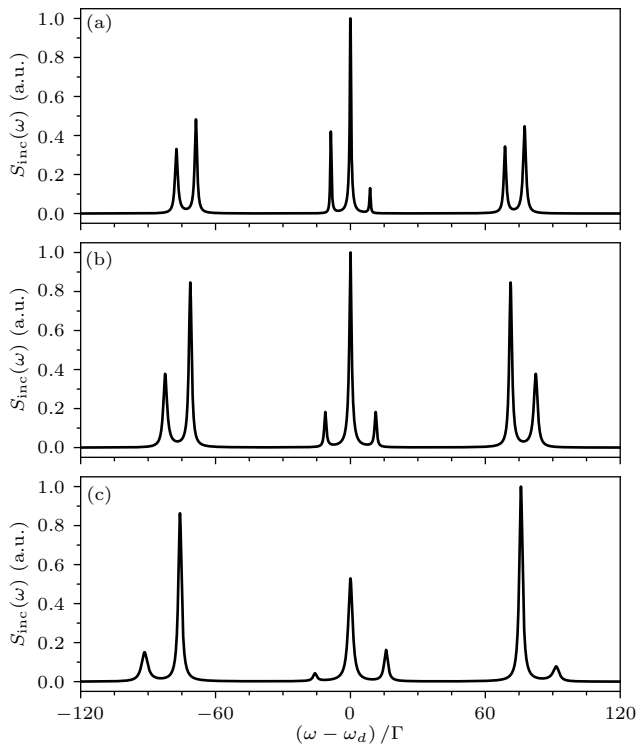}
  \caption{Normalized incoherent power spectrum at two-photon resonance $\left( \delta = 0 \right)$ for three different values of the dipole moment ratio: (a) $\xi = 1 / \sqrt{2}$, (b) $\xi = 1$, and (c) $\xi = \sqrt{2}$. The other parameters are $\Omega = 40 \Gamma$ and $\alpha = -120 \Gamma$.}
  \label{fig:3_spectrum_strong_drive_xi}
\end{figure}

As the driving amplitude increases and the degeneracy of the dressed state lifts, seven peaks emerge, grouped into three distinct structures \cite{Whitley_1976_PRA_DoubleOpticalResonance}. The two single-photon resonances, $\omega_{eg}$ and $\omega_{fe}$, are both shifted away from resonance and split into doublets due to the induced Stark shift \cite{Ardelt_2016_PRB_OpticalControlNonlinearly, Hargart_2016_PRB_CavityEnhancedSimultaneous, Bounouar_2017_PRL_PathControlledTime}. Meanwhile, a central peak emerges due to the two-photon driving. As the driving strength increases further, this central peak splits into a Mollow-like triplet as the three-level atom acts as an effective two-level atom from the resonant two-photon driving. We clearly observe the appearance of these spectral structures in Fig.~\ref{fig:3_spectrum_scan_Omega}, where the normalized incoherent power spectrum is plotted as a function of driving amplitude. We note that the separation between the peaks of the doublet and the separation of the peaks in the central triplet is exactly the dressed-state frequency $\omega_{l}$ which, in the strong driving limit, approaches the effective two-photon driving amplitude, Eq.~(\ref{eq:2_effective_two_photon_drive}).

As previously mentioned, these peaks, as labeled in Fig.~\ref{fig:3_dressed_states_labelled_spectrum}(c), correspond exactly to the seven dressed-state transition frequencies, Eq.~(\ref{eq:3_transition_frequencies}): $\tilde{\omega}_{0}$ [green in Fig.~\ref{fig:3_dressed_states_labelled_spectrum}(a)] locates the central peak, $\tilde{\omega}_{\pm 1}$ [blue and yellow in Fig.~\ref{fig:3_dressed_states_labelled_spectrum}(a)] locate the sidepeaks of the central triplet, $\tilde{\omega}_{\pm 2}$ [orange and purple in Fig.~\ref{fig:3_dressed_states_labelled_spectrum}(a)] locate the inner peaks of the side doublets, and $\tilde{\omega}_{\pm 3}$ [red and magenta in Fig.~\ref{fig:3_dressed_states_labelled_spectrum}(a)] locate the outer peaks of the side doublets.

The symmetry of the fluorescence spectrum is lost when $\xi \neq 1$, as shown in Fig.~\ref{fig:3_spectrum_strong_drive_xi}. As the dipole moment ratio increases, increasing the Stark shift of the $\ket{g}$ and $\ket{f}$ states, the side doublets move further out. As $\xi$ changes, the central triplet is also affected: for $\xi < 1$, the right peak of the central triplet is suppressed, and for $\xi > 1$, the left peak is suppressed. Holm and Sargent point to this asymmetry as being a result of the Stark shift introducing ``dispersive-like features'' \cite{Holm_1985_OL_TheoryTwoPhoton, Alexanian_2006_PRA_TwoPhotonResonance}.

\begin{figure}
  \centering
  {\fontsize{7pt}{1em}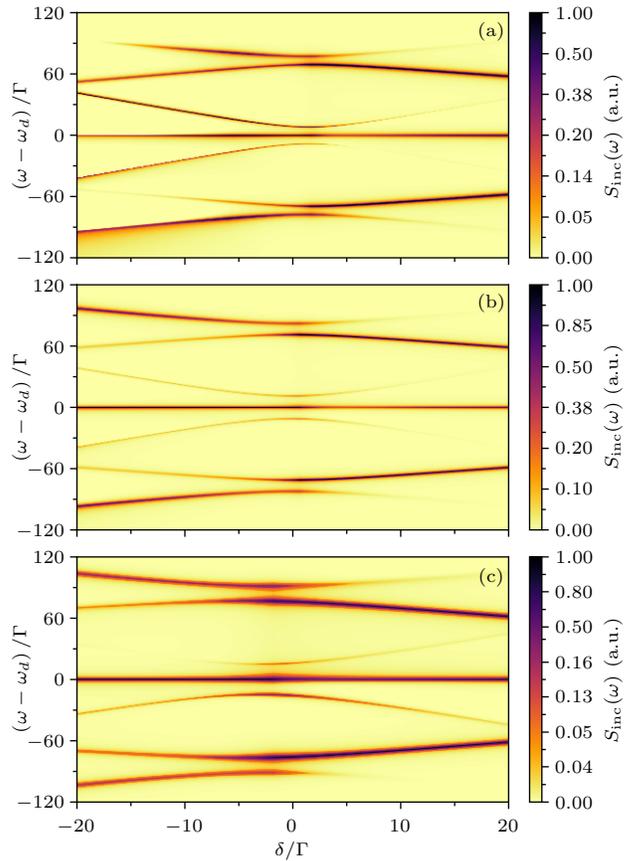}
  \caption{Normalized incoherent power spectrum as a function of drive detuning $\delta$ in the strong driving regime, $\Omega = 40 \Gamma$, for three different values of the dipole moment ratio: (a) $\xi = 1 / \sqrt{2}$, (b) $\xi = 1$, and (c) $\xi = \sqrt{2}$. The anharmonicity is $\alpha = -120 \Gamma$.}
  \label{fig:3_spectrum_scan_delta}
\end{figure}

In Fig.~\ref{fig:3_spectrum_scan_delta}, we stay in the strong driving regime, but move away from two-photon resonance, whereby we see a drastic change in the organization of the peaks. As the drive frequency shifts towards the single-photon resonance $\omega_{eg}$ -- positive values of $\delta$ for $\alpha < 0$ -- the three-level atom effectively behaves as a two-level atom. This is reflected in the fluorescence spectrum, where the seven-peak structure morphs into a three-peaked Mollow triplet. In the other direction -- for negative values of $\delta$ -- the side-doublet and central-triplet groupings are lost. The sidebands of the central triplet move outwards as the effective two-level approximation breaks down.

We also see the shift in the two-photon resonance frequencies for values of $\xi \neq 1$: Fig.~\ref{fig:3_spectrum_scan_delta}(a) shows a shift towards positive detunings for $\xi < 1$, and Fig.~\ref{fig:3_spectrum_scan_delta}(c) shows a shift towards negative detunings for $\xi > 1$.

These results align with previous works, most notably Ref.~\cite{Whitley_1976_PRA_DoubleOpticalResonance}. Whitley and Stroud consider the case where the two atomic transitions are driven independently, with two separate detunings $\delta_{a}$ and $\delta_{b}$ for the lower and upper transitions, respectively. In our model, both transitions are driven by a single, two-photon resonant field. Two-photon resonance $\delta = 0$ corresponds to $\delta_{a} = - \delta_{b}$, as in Fig.~7 of Ref.~\cite{Whitley_1976_PRA_DoubleOpticalResonance}, where the seven total emission peaks are visible in the two separate spectra. Figure~8 of Ref.~\cite{Whitley_1976_PRA_DoubleOpticalResonance}, however, depicts the fluorescence spectrum at single-photon resonance with the upper transition, corresponding to $\delta = \alpha / 2$ in our work.

\section{Atomic Photon Correlations}
\label{sect:4_atomic_g2}

\subsection{Total Radiated Field}

A standard, resonantly driven two-level atom can only store a single quantum of energy at any time. As the atom de-excites, and a photon is emitted, the atom must first be re-excited before it can emit another. The emitted light is therefore antibunched with an initial second-order correlation of $g^{(2)}(0) = 0$.

For the three-level ladder-type atom, we define the normalized second-order correlation function in the steady-state limit as:
\begin{equation}
  g^{(2)}_{ss}(\tau) = \frac{ \langle \Sigma_{+}(0) \Sigma_{+} \Sigma_{-}(\tau) \Sigma_{-}(0) \rangle }{ \langle \Sigma_{+} \Sigma_{-} \rangle_{ss}^{2} } .
\end{equation}
As the collective decay operator $\Sigma_{-}$, Eq.~(\ref{eq:2_atomic_raising_lowering_operators}), is a combination of the decay operators for the upper and lower dipole moments, the probability for an atomic decay to occur is proportional to $\Gamma \langle \Sigma_{+} \Sigma_{-} \rangle_{ss} = \Gamma \left( \langle \sigma_{ee}(t) \rangle + \xi^{2} \langle \sigma_{ff}(t) \rangle \right)$, which, we note, is dependent on the populations of both the intermediate and excited states.

\begin{figure}
  \centering
  {\fontsize{7pt}{0}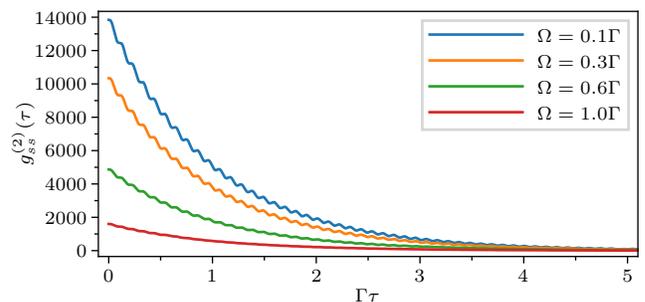}
  \caption{Atomic second-order correlation function in the weak-driving regime, with $\Omega = 0.1 \Gamma$ (blue), $0.3 \Gamma$ (orange), $0.6 \Gamma$ (green), and $\Gamma$ (red). The other parameters are $\alpha = -120 \Gamma, \delta = 0$, and $\xi = 1$.}
  \label{fig:4_g2_low_drive}
\end{figure}

In the low driving limit -- where the degeneracy of the dressed states has not yet been lifted -- the cascaded decay ensures that a photon of frequency $\omega_{fe}$ is always followed by a photon of frequency $\omega_{eg}$. This is characterized in the second-order correlation function as strong bunching, with high initial values, as seen in Fig.~\ref{fig:4_g2_low_drive}. As the driving amplitude decreases, the average rate at which the atom is excited also decreases, yet the average time between successive emissions remains the same. The initial correlation value grows as the cascaded emission becomes increasingly well-defined.

\begin{figure}
  \centering
  {\fontsize{7pt}{1em}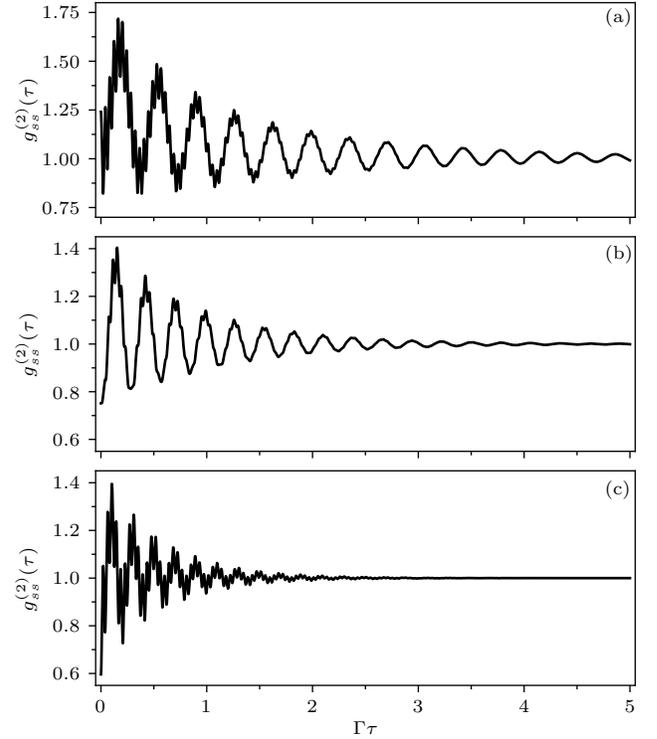}
  \caption{Atomic second-order correlation function in the strong driving regime, $\Omega = 40 \Gamma$, for three different values of the dipole moment ratio: (a) $\xi = 1 / \sqrt{2}$, (b) $\xi = 1$, and (c) $\xi = \sqrt{2}$. The other parameters are $\delta = 0$ and $\alpha = -120 \Gamma$.}
  \label{fig:4_g2_high_drive}
\end{figure}

Moving into the strong-driving regime in Fig.~\ref{fig:4_g2_high_drive}, we show correlation functions for three values of the dipole moment ratio: $\xi = 1 / \sqrt{2}$ [Fig.~\ref{fig:4_g2_high_drive}(a)], $\xi = 1$ [Fig.~\ref{fig:4_g2_high_drive}(b)], and $\xi = \sqrt{2}$ [Fig.~\ref{fig:4_g2_high_drive}(c)]. We see there are two distinct frequencies appearing in the correlations: the faster frequency is exactly the detuning from single-photon resonance, $\alpha / 2 + \delta$, while the slower frequency is the effective two-photon Rabi frequency, which is also the dressed-state frequency $\omega_{l}$. From Eqs.~(\ref{eq:2_effective_two_photon_drive}) and (\ref{eq:3_dressed_state_frequencies}), this slower frequency depends on $\xi$. This is evident in Fig.~\ref{fig:4_g2_high_drive}, where an increase in $\xi$ causes the correlation function to oscillate more rapidly, and also to decay more rapidly.

The dipole moment ratio also affects the initial value of the correlation function, $g^{(2)}_{ss}(0)$. From the effective two-level model, Eq.~(\ref{eq:2_approximate_steady_states}), the initial correlation value can be expressed as:
\begin{align}
  g^{(2)}_{ss}(0) &= \frac{ \langle \Sigma_{+}^{2} \Sigma_{-}^{2} \rangle_{ss} }{ \langle \Sigma_{+} \Sigma_{-} \rangle_{ss}^{2} } = \frac{ \Omega_{\mathrm{eff}}^{2} \left( 2 + \xi^{2} \right) + 4 \Delta_{\mathrm{eff}}^{2} + \xi^{4} \Gamma^{2} }{ 4 \xi^{2} \Omega_{\mathrm{eff}}^{4} } ,
\end{align}
which, at two-photon resonance $(\delta = 0)$, simplifies to:
\begin{equation}\label{eq:4_initial_g2_approximate}
  g^{(2)}_{ss}(0) = \frac{ 1 }{ 2 } + \frac{ 1 }{ 4 \xi^{2} } + \frac{ \alpha^{2} \Gamma^{2} }{ 4 \Omega^{4} } .
\end{equation}
Equation~(\ref{eq:4_initial_g2_approximate}) shows that, even as a two-level approximation, the photon correlations can never be perfectly antibunched and $g^{(2)}_{ss}(0) > 0.5$.

\subsection{Secular Approximation}

The dressed states and dressed-state frequencies in Eqs.~(\ref{eq:3_dressed_state_frequencies}) and (\ref{eq:3_dressed_states}) are derived only for two-photon resonance, where $\delta = 0$. They can, however, be calculated numerically for any detuning, given that
\begin{equation}
  \bm{D} = \bm{S}^{-1} H_{A} \bm{S},
\end{equation}
where $\bm{D} = \mathrm{diag} \left[ \omega_{m}, \omega_{u}, \omega_{l} \right]$ is a diagonal matrix of the eigenvalues of $H_{A}$, and $\bm{S} = \left[ \bm{v}_{m}, \bm{v}_{u}, \bm{v}_{l} \right]$ is a matrix composed of the three corresponding eigenvectors. We can then transform the master equation, Eq.~(\ref{eq:2_atomic_master_equation}), into the ``dressed-state basis'', with
\begin{equation}\label{eq:4_master_equation_dressed_state}
  \ddt \rho_{D} = \ddt \bm{S}^{-1} \rho \bm{S} = \frac{ 1 }{ i \hbar } \bm{S}^{-1} [H_{A}, \rho] \bm{S} + \frac{ \Gamma }{ 2 } \bm{S}^{-1} \Lambda(\Sigma_{-}) \rho \bm{S},
\end{equation}
where $\rho_{D} = \bm{S}^{-1} \rho \bm{S}$ is the atomic density operator in the dressed-state basis. We take a rotating wave approximation in the strong driving regime ($\Omega \gg \Gamma$) and drop any rapidly oscillating terms, such that, for $\delta = 0$, the atomic master equation in the dressed-state basis approximation is \cite{Whitley_1976_PRA_DoubleOpticalResonance, Narducci_1990_PRA_SpontaneousEmissionAbsorption}:
\begin{align}\label{eq:4_dressed_state_master_equation}
  \ddt[\rho_{D}] &= \frac{ 1 }{ i \hbar } [ H_{A} , \rho_{D} ] + \frac{ \Gamma_{0} }{ 2 } \Lambda \left( \sigma_{0} \right) \rho_{D} \nonumber \\
  &\quad +  \sum_{i=1,2,3} \frac{ \Gamma_{i} }{ 2 } \left[ \Lambda \left( \sigma_{+i} \right) \rho_{D} + \Lambda \left( \sigma_{-i} \right) \rho_{D} \right] ,
\end{align}
where $[ \sigma_{+i} = ( \sigma_{-i} )^{\dagger} ]$
\begin{subequations}\label{eq:4_3LA_dressed_state_operators}
  \begin{align}
    \sigma_{0} &= \ket{u} \bra{u} - \ket{l} \bra{l} , \label{eq:4_3LA_dressed_state_operators_0} \\
    \sigma_{+1} &= \ket{l} \bra{m} , \\
    \sigma_{+2} &= \ket{m} \bra{u} , \\
    \sigma_{+3} &= \ket{l} \bra{u} ,
  \end{align}
\end{subequations}
are the dressed-state operators corresponding to each of the dressed-state transition frequencies appearing in the fluorescence spectrum, Eq.~(\ref{eq:3_transition_frequencies}).

In general, the decay rates $\Gamma_{0}, \Gamma_{1}, \Gamma_{2}$, and $\Gamma_{3}$ are complicated expressions. However, in the limit $\Omega \rightarrow \infty$, these expressions reduce to
\begin{subequations}
  \begin{alignat}{3}
    \Gamma_{0} &\equiv \frac{ \left( 1 + \xi^{2} \right) \Gamma }{ 4 }, &\quad \Gamma_{1} &\equiv \frac{ \xi^{2} \Gamma }{ 2 \left( 1 + \xi^{2} \right) } , \\
    \Gamma_{2} &\equiv \frac{ \xi^{2} \Gamma }{ 2 \left( 1 + \xi^{2} \right) } , &\quad \Gamma_{3} &\equiv \frac{ \left( 1 - \xi^{2} \right)^{2} \Gamma }{ 4 \left( 1 + \xi^{2} \right) } .
  \end{alignat}
\end{subequations}
We then recast Eq.~(\ref{eq:4_dressed_state_master_equation}) into a set of moment equations for the dressed-state operators:
\begin{subequations}
  \begin{align}
    \ddt \langle \sigma_{\pm 1} \rangle &= -\left( \frac{ \Gamma_{1} }{ 2 } + i \omega_{\pm 1} \right) \langle \sigma_{\pm 1} \rangle , \\
    \ddt \langle \sigma_{\pm 2} \rangle &= -\left( \frac{ \Gamma_{2} }{ 2 } + i \omega_{\pm 2} \right) \langle \sigma_{\pm 2} \rangle , \\
    \ddt \langle \sigma_{\pm 3} \rangle &= -\left( \frac{ \Gamma_{3} }{ 2 } + i \omega_{\pm 3} \right) \langle \sigma_{\pm 3} \rangle ,
  \end{align}
\end{subequations}
and
\begin{equation}\label{eq:4_dressed_state_3x3_moment_equation}
  \ddt
  \begin{pmatrix}
    \langle \sigma_{mm} \rangle \\
    \langle \sigma_{uu} \rangle \\
    \langle \sigma_{ll} \rangle
  \end{pmatrix}
  = \bm{M}
  \begin{pmatrix}
    \langle \sigma_{mm} \rangle \\
    \langle \sigma_{uu} \rangle \\
    \langle \sigma_{ll} \rangle
  \end{pmatrix} ,
\end{equation}
with $\sigma_{ii} = \ket{i} \bra{i}$ and
\begin{equation}
  \bm{M} =
  \begin{psmallmatrix}
    -\Gamma \left( a_{4}^{2} + a_{7}^{2} \right) & \Gamma a_{2}^{2} & \Gamma a_{3}^{2} \\
    \Gamma a_{4}^{2} & -\Gamma \left( a_{2}^{2} + a_{8}^{2} \right) & \Gamma a_{6}^{2} \\
    \Gamma a_{7}^{2} & \Gamma a_{8}^{2} & -\Gamma \left( a_{3}^{2} + a_{6}^{2} \right)
  \end{psmallmatrix} ,
\end{equation}
where $\{ a_{i} \}$ are defined in Appendix~\ref{app:2_dressed_state_derivation}.

The generalized second-order correlation function for the dressed-state transitions in the steady-state limit can then be written as $(A, B = 0, \pm 1, \pm 2, \pm 3)$:
\begin{equation}
  g^{(2)}(\omega_{A}, 0; \omega_{B}, \tau) = \frac{ \langle \sigma_{A}^{\dagger}(0) \sigma_{B}^{\dagger} \sigma_{B}(\tau) \sigma_{A}(0) \rangle }{ \langle \sigma_{A}^{\dagger} \sigma_{A} \rangle_{ss} \langle \sigma_{B}^{\dagger} \sigma_{B} \rangle_{ss} } .
\end{equation}
Using the quantum regression equations \cite{Carmichael2002}, we can derive explicit expressions for the two-time second-order correlation functions from the general solution to Eq.~(\ref{eq:4_dressed_state_3x3_moment_equation}):
\begin{subequations}\label{eq:4_general_solution_two_time_correlation}
  \begin{align}
    \langle \sigma_{A}^{\dagger}(0) \sigma_{mm}(\tau) \sigma_{A}(0) \rangle &= C_{1} - 2 C_{2} e^{\lambda_{-} \tau} , \\
    \langle \sigma_{A}^{\dagger}(0) \sigma_{uu}(\tau) \sigma_{A}(0) \rangle &= C_{1} + C_{2} e^{\lambda_{-} \tau} - C_{3} e^{\lambda_{+} \tau} , \\
    \langle \sigma_{A}^{\dagger}(0) \sigma_{ll}(\tau) \sigma_{A}(0) \rangle &= C_{1} + C_{2} e^{\lambda_{-} \tau} + C_{3} e^{\lambda_{+} \tau} ,
  \end{align}
\end{subequations}
with
\begin{subequations}
  \begin{align}
    C_{1} &\equiv \frac{1}{3} \Big( \langle \sigma_{A}^{\dagger} \sigma_{ll} \sigma_{A} \rangle_{ss} + \langle \sigma_{A}^{\dagger} \sigma_{uu} \sigma_{A} \rangle_{ss} + \langle \sigma_{A}^{\dagger} \sigma_{mm} \sigma_{A} \rangle_{ss} \Big) , \\
    C_{2} &\equiv \frac{1}{6} \Big( \langle \sigma_{A}^{\dagger} \sigma_{ll} \sigma_{A} \rangle_{ss} + \langle \sigma_{A}^{\dagger} \sigma_{uu} \sigma_{A} \rangle_{ss} - 2 \langle \sigma_{A}^{\dagger} \sigma_{mm} \sigma_{A} \rangle_{ss} \Big) , \\
    C_{3} &\equiv \frac{1}{2} \Big( \langle \sigma_{A}^{\dagger} \sigma_{ll} \sigma_{A} \rangle_{ss} - \langle \sigma_{A}^{\dagger} \sigma_{uu} \sigma_{A} \rangle_{ss} \Big) ,
  \end{align}
\end{subequations}
and
\begin{subequations}
  \begin{align}
    \lambda_{-} &\equiv \frac{ -3 \xi^{2} \Gamma }{ 2 \left( 1 + \xi^{2} \right) } , \\
    \lambda_{+} &\equiv \frac{ -\Gamma \left( 1 - \xi^{2} + \xi^{4} \right) }{ 2 \left( 1 + \xi^{2} \right) } .
  \end{align}
\end{subequations}

For more details on the moment equation derivation and generalized expressions for the decay rates and evolution matrix, see Appendix \ref{app:2_dressed_state_derivation}.

\subsubsection{Auto-correlations}

\begin{figure}
  \centering
  {\fontsize{7pt}{1em}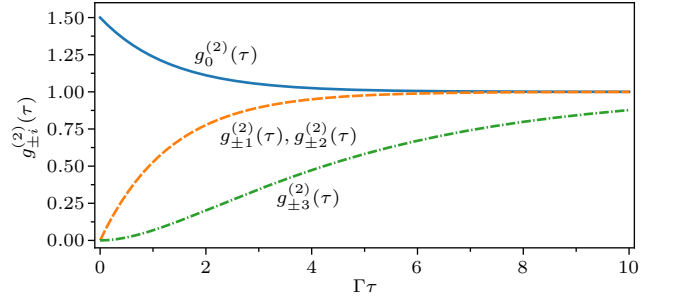}
  \caption{Auto-correlation functions, Eqs.~(\ref{eq:4_dressed_state_auto_correlations}), of the central peak (blue, solid), $\tilde{\omega}_{\pm 1}$ and $\tilde{\omega}_{\pm 2}$ sidepeaks (orange, dashed), and $\tilde{\omega}_{\pm 3}$ sidepeaks (green, dash-dot). The dipole moment ratio is $\xi = 1$.}
  \label{fig:4_dressed_state_g2_auto}
\end{figure}

We first consider the auto-correlations of each of the dressed-state transitions, with  $g^{(2)}_{\pm i}(\tau) = g^{(2)}(\omega_{\pm i}, 0: \omega_{\pm i}, \tau)$. From the general solution Eq.~(\ref{eq:4_general_solution_two_time_correlation}), the auto-correlation functions of each of the dressed-state transitions take the straightforward forms,
\begin{subequations}\label{eq:4_dressed_state_auto_correlations}
  \begin{gather}
    g^{(2)}_{0}(\tau) = 1 + \frac{ 1 }{ 2 } e^{\lambda_{-} \tau} , \label{eq:4_g2_auto_dressed_0} \\
    g^{(2)}_{\pm 1}(\tau) = 1 - e^{\lambda_{-} \tau} , \label{eq:4_g2_auto_dressed_1} \\
    g^{(2)}_{\pm 2}(\tau) = 1 - e^{\lambda_{-} \tau} , \label{eq:4_g2_auto_dressed_2} \\
    g^{(2)}_{\pm 3}(\tau) = 1 + \frac{ 1 }{ 2 } e^{\lambda_{-} \tau} - \frac{ 3 }{ 2 }  e^{\lambda_{+} \tau} . \label{eq:4_g2_auto_dressed_3}
  \end{gather}
\end{subequations}
These auto-correlation functions, as shown in Fig.~\ref{fig:4_dressed_state_g2_auto}, show agreement with the transitions depicted in the energy level diagram in Fig.~\ref{fig:3_dressed_states_labelled_spectrum}(a). Each of the sidepeaks emissions are antibunched, with an initial value of $g^{(2)}(0) = 0$. Following the emission of one photon, the atomic population is shifted into a different dressed state. Therefore, the atom must be re-excited before it can emit another photon of the same frequency. The central peak, however, exhibits slight bunching, with an initial value of $g^{(2)}(0) = 1.5$. While we expect the $\ket{u} \rightarrow \ket{u}$ and $\ket{l} \rightarrow \ket{l}$ transitions to be bunched in the two-photon decay due to the cascaded emissions, there are two degenerate, uncorrelated transitions. The overall bunching of the central peak is therefore diminished.

Note that the auto-correlation functions for the $\tilde{\omega}_{\pm 1}$ and $\tilde{\omega}_{\pm 2}$ peaks are identical. In the limit $\Omega \rightarrow \infty$, the $\ket{u}$ and $\ket{l}$ dressed-state frequencies have opposite signs, $\omega_{u} \approx -\omega_{l}$, hence the $\tilde{\omega}_{\pm 1}$ and $\tilde{\omega}_{\mp 2}$ transition frequencies are identical \cite{Whitley_1976_PRA_DoubleOpticalResonance}.

\subsubsection{Cross-correlations}

There are seven distinct frequency components in the fluorescence spectrum, with a total of 49 possible combinations of cross-correlations. While some of these may be identical -- seven of these are, of course, the auto-correlations Eqs.~(\ref{eq:4_dressed_state_auto_correlations}) -- there are still too many to practically list. However, the general solution, Eq.~(\ref{eq:4_general_solution_two_time_correlation}), allows us to calculate exact expressions for the various cross-correlation functions. Therefore, in this section, we present only a select few.

\begin{figure}
  \centering
  {\fontsize{7pt}{1em}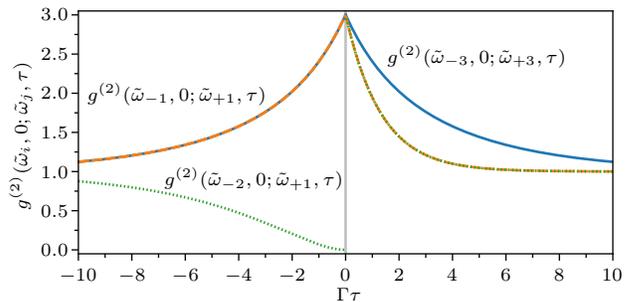}
  \caption{Cross-correlation functions, Eqs.~(\ref{eq:4_dressed_state_cross_correlations})-(\ref{eq:4_cross_correlations_not_opposite_peaks}), in forwards and backwards time for the following transitions: $\tilde{\omega}_{-3} \rightarrow \tilde{\omega}_{+3}$ (blue, solid); $\tilde{\omega}_{-1} \rightarrow \tilde{\omega}_{+1}$ (orange, dashed); and $\tilde{\omega}_{+2} \rightarrow \tilde{\omega}_{+1}$ (green, dotted). The dipole moment ratio is $\xi = 1$.}
  \label{fig:4_dressed_state_g2_cross}
\end{figure}

The ``natural'' cross-correlations to compute are those between opposite sidepeaks, e.g., $\tilde{\omega}_{-1}$ followed by $\tilde{\omega}_{+1}$. For the three different sidepeaks, the cross-correlation functions are:
\begin{subequations}\label{eq:4_dressed_state_cross_correlations}
  \begin{align}
    g^{(2)}(\tilde{\omega}_{-1}, 0; \tilde{\omega}_{+1}, \tau) &= 1 + 2 e^{\lambda_{-} \tau} \label{eq:4_cross_corelation_-1_+1}, \\
    g^{(2)}(\tilde{\omega}_{-2}, 0; \tilde{\omega}_{+2}, \tau) &= 1 + 2 e^{\lambda_{-} \tau} , \\
    g^{(2)}(\tilde{\omega}_{-3}, 0; \tilde{\omega}_{+3}, \tau) &= 1 + \frac{ 1 }{ 2 } e^{\lambda_{-} \tau} + \frac{ 3 }{ 2 } e^{\lambda_{+} \tau} .
  \end{align}
\end{subequations}
In the opposite ordering, $\tilde{\omega}_{+i}$ followed by $\tilde{\omega}_{-1}$, the expressions are identical, with
\begin{equation}\label{eq:4_dressed_state_cross_correlations_opposite_order}
  g^{(2)}(\tilde{\omega}_{+i}, 0; \tilde{\omega}_{-i}, \tau) = 1 + \frac{ 1 }{ 2 } e^{\lambda_{-} \tau} + \frac{ 3 }{ 2 } e^{\lambda_{+} \tau} ,
\end{equation}
as can be seen in Fig.~\ref{fig:4_dressed_state_g2_cross}. These correlation functions display bunching in forwards and backwards time, with an initial value of $g^{(2)}(0) = 3$, as the first emission puts the atom into the corresponding initial state for the second emission.

Following the decay paths in Fig.~\ref{fig:3_dressed_states_labelled_spectrum}(a), there are some transitions one might expect to be correlated in one direction and anti-correlated in the other, e.g., correlating a photon of frequency $\tilde{\omega}_{+2}$ with a photon of frequency $\tilde{\omega}_{+1}$. For an emission of an $\tilde{\omega}_{+2}$ photon, the atom decays via the transition $\ket{u} \rightarrow \ket{m}$, whereas for an emission of an $\tilde{\omega}_{+1}$ photon, the atom decays via $\ket{m} \rightarrow \ket{l}$. Given a detection of an $\tilde{\omega}_{+2}$ photon first, the atom is now in the correct initial state for an $\tilde{\omega}_{+1}$ emission, $\ket{m}$; therefore we expect these two transitions to be correlated. In the opposite direction, however, the $\tilde{\omega}_{+1}$ emission puts the atom into the $\ket{l}$ state; hence the atom cannot emit an $\tilde{\omega}_{+2}$ photon and the two transitions are anti-correlated. This is reflected in the cross-correlation functions:
\begin{subequations}\label{eq:4_cross_correlations_not_opposite_peaks}
  \begin{align}
    g^{(2)}(\tilde{\omega}_{+2}, 0; \tilde{\omega}_{+1}, \tau) &= 1 + 2 e^{\lambda_{-} \tau} , \\
    g^{(2)}(\tilde{\omega}_{+1}, 0; \tilde{\omega}_{+2}, \tau) &= 1 + \frac{ 1 }{ 2 } e^{\lambda_{-} \tau} - \frac{ 3 }{ 2 } e^{\lambda_{-} \tau} ,
  \end{align}
\end{subequations}
which are also depicted in Fig.~\ref{fig:4_dressed_state_g2_cross} (green, dotted curve).

\section{Conclusion}

In this work, we have presented numerical and analytic results for the steady-state behavior, fluorescence spectrum, and photon correlations of a driven three-level ladder-type atom. Previous work -- Refs.~\cite{Whitley_1976_PRA_DoubleOpticalResonance} and \cite{Manka_1991_PRA_SpontaneousEmissionAbsorption}, in particular -- have calculated emission and absorption spectra for generalized three-level ladder-type atoms, yet here we have extended the theory to a specific configuration: namely, an anharmonic ladder-type atom driven at two-photon resonance. Given a large anharmonicity between the upper and lower dipoles, the atom must be excited to its upper state via the simultaneous absorption of two photons. The atom then de-excites back to the ground state via a two-photon cascaded process.

When strongly driven with a coherent, two-photon resonant field, the fluorescence spectrum exhibits seven peaks, neatly organized into three distinct groups: a central, Mollow-like triplet from the effective two-photon drive, and two side-doublets from the Stark splitting of the bare states.

The cascaded ordering of the de-excitation also gives rise to strong bunching in the photon correlations. When weakly driven, the atom will emit a photon from the upper dipole and then from the lower dipole. As the driving amplitude gets weaker and the cascaded emissions become increasingly well-defined, the initial correlation value grows. At strong driving amplitudes, the three-level atom behaves like a two-level atom; the photon correlations have an additional oscillation at the effective Rabi frequency. Depending on the dipole moment ratio, the emitted light is also slightly antibunched, with $0.5 < g^{(2)}(0) < 1$, though it is never perfectly antibunched.

Taking inspiration from previous work on the two-level atom -- Ref.~\cite{Schrama_1992_PRA_IntensityCorrelationsComponents} in particular -- we have derived idealized expressions for frequency-resolved photon correlations. In the secular approximation, the behavior of the dressed-state auto- and cross-correlations agree with the cascaded structure of the dressed-state de-excitation pathways. Of course, the secular approximation is still an approximation. In a separate work \cite{Ngaha_PhD_Thesis_2023}, we have taken this a step further by modeling the case in which the fluorescence is directed into cavity-based frequency-filters, whereby we can correlate the outputs of the frequency-filters to compare against the secular approximation \cite{Ngaha_2024_PRA_MultimodeArrayFiltering}.

\appendix

\section{Effective Two-Level Drive}
\label{app:1_effective_2LA_model}

We begin by solving the time-dependent Schr{\"{o}}dinger equation for the atomic state vector with the Hamiltonian Eq.~(\ref{eq:2_H_atom_time_indepdent}), where, for the pure state
\begin{equation}
  \ket{\psi} = c_{g} \ket{g} + c_{e} \ket{e} + c_{f} \ket{f},
\end{equation}
we obtain three coupled ordinary differential equations:
\begin{subequations}\label{eq:2_atomic_schrodinger}
  \begin{align}
    \dot{c}_{g} &= -i \frac{ \Omega }{ 2 } c_{e} , \\
    \dot{c}_{e} &= -i \frac{ \Omega }{ 2 } c_{g} + i \left( \frac{ \alpha }{ 2 } + \delta \right) c_{e} - i \xi \frac{ \Omega }{ 2 } c_{f} , \\
    \dot{c}_{f} &= -i \xi \frac{ \Omega }{ 2 } c_{e} + 2 i \delta c_{f}.
  \end{align}
\end{subequations}
We set $\dot{c}_{e} = 0$ and find the quasi-steady state for the intermediate state amplitude:
\begin{equation}
  c_{e}^{ss} = \frac{ \Omega }{ 2 } \frac{ c_{g} + \xi c_{f} }{ \alpha / 2 + \delta } .
\end{equation}
Equations~(\ref{eq:2_atomic_schrodinger}) then reduce to a pair of coupled equations for the ground and excited states:
\begin{subequations}
  \begin{align}
    \dot{c}_{g} &= -i \left( \frac{ \Omega }{ 2 } \right)^{2} \frac{ c_{g} }{ \alpha / 2 + \delta } - i \xi \left( \frac{ \Omega }{ 2 } \right)^{2} \frac{ c_{f} }{ \alpha / 2 + \delta } , \\
    \dot{c}_{f} &= -i \xi \left( \frac{ \Omega }{ 2 } \right)^{2} \frac{ c_{g} }{ \alpha / 2 + \delta } -i \left[ \left( \frac{ \xi \Omega }{ 2 } \right)^{2} \frac{ 1 }{ \alpha / 2 + \delta } - 2 \delta \right] c_{f} .
  \end{align}
\end{subequations}
From these equations, we define the effective two-photon driving amplitude and the level shifts of the ground and excited states:
\begin{subequations}
  \begin{align}
    \frac{ \Omega_{\mathrm{eff}} }{ 2 } &\equiv \xi \left( \frac{ \Omega }{ 2 } \right)^{2} \frac{ 1 }{ \alpha / 2  + \delta }, \\
    \Delta_{g} &\equiv \left( \frac{ \Omega }{ 2 } \right)^{2} \frac{ 1 }{ \alpha / 2 + \delta }, \\
    \Delta_{f} &\equiv \left( \frac{ \xi \Omega }{ 2 } \right)^{2} \frac{ 1 }{ \alpha / 2 + \delta } .
  \end{align}
\end{subequations}
We therefore define the \textit{effective two-photon drive Hamiltonian} as:
\begin{align}
  H_{\mathrm{eff}} &= \hbar \Delta_{g} \ket{g} \bra{g} + \hbar \left( -2 \delta + \Delta_{f} \right) \ket{f} \bra{f} \nonumber \\
  &\quad + \hbar \frac{ \Omega_{\mathrm{eff}} }{ 2 } \Big( \ket{g} \bra{f} + \ket{f} \bra{g} \Big) .
\end{align}

Assuming a large atomic anharmonicity, $| \alpha | \gg \Gamma$, we can drop any rapidly oscillating cross terms in the master equation, Eq.~(\ref{eq:2_atomic_master_equation}), such that
\begin{equation}
  \ddt[\rho] = \frac{ 1 }{ i \hbar } [H_{\mathrm{eff}}, \rho ] + \frac{ \Gamma }{ 2 } \Lambda(\sigma_{ge}) \rho + \frac{ \xi^{2} \Gamma }{ 2 } \Lambda(\sigma_{ef}) \rho .
\end{equation}
From this, we can derive a complete set of operator moment equations for each dipole raising and lowering operator $\left( \sigma_{ij} = \ket{i} \bra{j} \right)$:
\begin{subequations}\label{eq:A_main_moment_equations}
  \begin{align}
    \ddt \langle \sigma_{gg} \rangle &= \Gamma \langle \sigma_{ee} \rangle - i \frac{ \Omega_{\mathrm{eff}} }{ 2 } \langle \sigma_{gf} \rangle + i \frac{ \Omega_{\mathrm{eff}} }{ 2 } \langle \sigma_{fg} \rangle , \\
    \ddt \langle \sigma_{ee} \rangle &= -\Gamma \langle \sigma_{ee} \rangle + \xi^{2} \Gamma \langle \sigma_{ff} \rangle , \\
    \ddt \langle \sigma_{ff} \rangle &= -\xi^{2} \Gamma \langle \sigma_{ff} \rangle + i \frac{ \Omega_{\mathrm{eff}} }{ 2 } \langle \sigma_{gf} \rangle - i \frac{ \Omega_{\mathrm{eff}} }{ 2 } \langle \sigma_{fg} \rangle, \\
    \ddt \langle \sigma_{gf} \rangle &= -\left( \frac{ \xi^{2} \Gamma }{ 2 } + i \Delta_{\mathrm{eff}} \right) \langle \sigma_{gf} \rangle - i \frac{ \Omega_{\mathrm{eff}} }{ 2 } \langle \sigma_{gg} \rangle \\ \nonumber
    &\quad + i \frac{ \Omega_{\mathrm{eff}} }{ 2 } \langle \sigma_{ff} \rangle, \\
    \ddt \langle \sigma_{fg} \rangle &= -\left( \frac{ \xi^{2} \Gamma }{ 2 } - i \Delta_{\mathrm{eff}} \right) \langle \sigma_{fg} \rangle + i \frac{ \Omega_{\mathrm{eff}} }{ 2 } \langle \sigma_{gg} \rangle \nonumber \\
    &\quad - i \frac{ \Omega_{\mathrm{eff}} }{ 2 } \langle \sigma_{ff} \rangle ,
  \end{align}
\end{subequations}
and
\begin{subequations}
  \begin{align}
    \ddt \langle \sigma_{ge} \rangle &= -\left( \frac{ \Gamma }{ 2 } - i \Delta_{g} \right) \langle \sigma_{ge} \rangle + i \frac{ \Omega_{\mathrm{eff}} }{ 2 } \langle \sigma_{fe} \rangle , \\
    \ddt \langle \sigma_{eg} \rangle &= -\left( \frac{ \Gamma }{ 2 } + i \Delta_{g} \right) \langle \sigma_{eg} \rangle - i \frac{ \Omega_{\mathrm{eff}} }{ 2 } \langle \sigma_{ef} \rangle , \\
    \ddt \langle \sigma_{ef} \rangle &= -\left[ \frac{ \Gamma }{ 2 } \left( 1 + \xi^{2} \right) - i \left( 2\delta - \Delta_{f} \right) \right] \langle \sigma_{ef} \rangle \nonumber \\
    &\quad - i \frac{ \Omega_{\mathrm{eff}} }{ 2 } \langle \sigma_{eg} \rangle, \\
    \ddt \langle \sigma_{fe} \rangle &= -\left[ \frac{ \Gamma }{ 2 } \left( 1 + \xi^{2} \right) + i \left( 2\delta - \Delta_{f} \right) \right] \langle \sigma_{ef} \rangle \nonumber \\
    &\quad + i \frac{ \Omega_{\mathrm{eff}} }{ 2 } \langle \sigma_{ge} \rangle .
  \end{align}
\end{subequations}
Using the fact that $\langle \sigma_{gg} \rangle + \langle \sigma_{ee} \rangle + \langle \sigma_{ff} \rangle = 1$, Eqs.~(\ref{eq:A_main_moment_equations}) further reduce to a set of four coupled equations \cite{Whitley_1976_PRA_DoubleOpticalResonance}:
\begin{subequations}
  \begin{align}
    \ddt \langle \sigma_{gg} \rangle &= -\Gamma \langle \sigma_{gg} \rangle - \Gamma \langle \sigma_{ff} \rangle - i \frac{ \Omega_{\mathrm{eff}} }{ 2 } \langle \sigma_{gf} \rangle \nonumber \\
    &\quad + i \frac{ \Omega_{\mathrm{eff}} }{ 2 } \langle \sigma_{fg} \rangle + \Gamma , \\
    \ddt \langle \sigma_{ff} \rangle &= -\xi^{2} \Gamma \langle \sigma_{ff} \rangle + i \frac{ \Omega_{\mathrm{eff}} }{ 2 } \langle \sigma_{gf} \rangle - i \frac{ \Omega_{\mathrm{eff}} }{ 2 } \langle \sigma_{fg} \rangle, \\
    \ddt \langle \sigma_{gf} \rangle &= -i \frac{ \Omega_{\mathrm{eff}} }{ 2 } \langle \sigma_{gg} \rangle + i \frac{ \Omega_{\mathrm{eff}} }{ 2 } \langle \sigma_{ff} \rangle \nonumber \\
    &\quad - \left( \frac{ \xi^{2} \Gamma }{ 2 } + i \Delta_{\mathrm{eff}} \right) \langle \sigma_{gf} \rangle , \\
    \ddt \langle \sigma_{fg} \rangle &= i \frac{ \Omega_{\mathrm{eff}} }{ 2 } \langle \sigma_{gg} \rangle - i \frac{ \Omega_{\mathrm{eff}} }{ 2 } \langle \sigma_{ff} \rangle \nonumber \\
    &\quad - \left( \frac{ \xi^{2} \Gamma }{ 2 } - i \Delta_{\mathrm{eff}} \right) \langle \sigma_{fg} \rangle .
  \end{align}
\end{subequations}

\section{Dressed-State Approximation}
\label{app:2_dressed_state_derivation}

Writing the bare atomic states as unit vectors,
\begin{equation}
  \ket{g} \rightarrow
  \begin{pmatrix}
    1 \\
    0 \\
    0
  \end{pmatrix}
  , \quad \ket{e} \rightarrow
  \begin{pmatrix}
    0 \\
    1 \\
    0
  \end{pmatrix}
  , \quad \ket{f} \rightarrow
  \begin{pmatrix}
    0 \\
    0 \\
    1
  \end{pmatrix},
\end{equation}
we can also represent Hamiltonian, Eq.~(\ref{eq:2_H_atom_time_indepdent}), in matrix form:
\begin{equation}
  H_{A} \rightarrow \hbar
  \begin{pmatrix}
    0 & \frac{ \Omega }{ 2 } & 0 \\
    \frac{ \Omega }{ 2 } & -\left( \frac{ \alpha }{ 2 } + \delta \right) & \xi \frac{ \Omega }{ 2 } \\
    0 & \xi \frac{ \Omega }{ 2 } & -2 \delta
  \end{pmatrix} .
\end{equation}
This matrix is diagonalizable, with
\begin{equation}
  \bm{D} = \bm{S}^{-1} H_{A} \bm{S},
\end{equation}
where
\begin{equation}
  \bm{D} = \mathrm{diag} \left[ \omega_{m}, \omega_{u}, \omega_{l} \right],
\end{equation}
is a diagonal matrix consisting of the eigenvalues, with $\omega_{l} \leq \omega_{m} \leq \omega_{u}$, and
\begin{equation}
  \bm{S} = \left[ \ket{m} , \ket{u} , \ket{l} \right]
\end{equation}
is a matrix consisting of the corresponding eigenvectors written in the bare-state basis. The inverse of this matrix, $\bm{S}^{-1}$, then gives the bare atomic states in the ``dressed-state basis''.

We can transform the atomic lowering operators Eq.~(\ref{eq:2_atomic_raising_lowering_operators}), into the dressed-state basis with
\begin{align}
  {\Sigma}_{-}^{\prime} = \bm{S}^{-1} \Sigma_{-} \bm{S} &= \bm{S}^{-1} \Big( \ket{g} \bra{e} + \xi  \ket{e} \bra{f} \Big) \bm{S} .
\end{align}
Due to the cubic nature of the characteristic polynomial, generalized analytic expressions for the dressed states for non-zero detuning are exceedingly difficult to derive. As eigenvectors can be easily computed numerically, we introduce a change of variable by defining the elements of the atomic lowering operator in the dressed-state basis:
\begin{equation}
  \Sigma_{-}^{\prime} \longrightarrow
  \begin{pmatrix}
    \ket{m} \\
    \ket{u} \\
    \ket{l}
  \end{pmatrix}
  \begin{pmatrix}
    a_{1} & a_{2} & a_{3} \\
    a_{4} & a_{5} & a_{6} \\
    a_{7} & a_{8} & a_{9}
  \end{pmatrix}
  \begin{pmatrix}
    \ket{m} & \ket{u} & \ket{l}
  \end{pmatrix},
\end{equation}
with which we will derive generalized expressions for the secular approximation.

Transforming into an interaction picture with the unitary evolution operator
\begin{equation}
  \mathcal{U}(t) = e^{\frac{ 1 }{ i \hbar } H_{A} t},
\end{equation}
the atomic lowering operator then evolves as:
\begin{align}
  \tilde{\Sigma}_{-}^{\prime}(t) &= \mathcal{U}^{\dagger}(t) \Sigma_{-}^{\prime} ~ \mathcal{U}(t) \nonumber \\
  &\rightarrow
  \begin{pmatrix}
    a_{1} & a_{2} e^{-i \left( \omega_{u} - \omega_{m} \right) t} & a_{3} e^{i \left( \omega_{m} - \omega_{l} \right) t} \\
    a_{4} e^{i \left( \omega_{u} - \omega_{m} \right) t} & a_{5} & a_{6} e^{i \left( \omega_{u} - \omega_{l} \right) t} \\
    a_{7} e^{-i \left( \omega_{m} - \omega_{l} \right) t} & a_{8} e^{-i \left( \omega_{u} - \omega_{l} \right) t} & a_{9}
  \end{pmatrix} .
\end{align}
The atomic master equation, Eq.~(\ref{eq:2_H_atom_time_indepdent}), also transforms as:
\begin{equation}\label{eq:B_dressed_state_master_equation_interaction_picture}
  \ddt[\tilde{\rho}_{D}] = \frac{ \Gamma }{ 2 } \Lambda \left( \tilde{\Sigma}_{-}^{\prime}(t) \right) \tilde{\rho}_{D}(t),
\end{equation}
where
\begin{equation}
  \tilde{\rho}_{D}(t) = \mathcal{U}^{\dagger}(t) \rho_{D} ~ \mathcal{U}(t), \quad \rho_{D} = \bm{S}^{-1} \rho ~ \bm{S} .
\end{equation}
Expanding the operators in Eq.~(\ref{eq:B_dressed_state_master_equation_interaction_picture}), we then make the rotating wave approximation and drop any rapidly oscillating terms. Transforming back out of the interaction picture, the master equation can be rewritten in terms of dressed-state operators \cite{Whitley_1976_PRA_DoubleOpticalResonance, Narducci_1990_PRA_SpontaneousEmissionAbsorption}:
\begin{align}
  \ddt[\rho_{D}] &= -i [\omega_{m} \sigma_{mm} + \omega_{u} \sigma_{uu} + \omega_{l} \sigma_{ll}, \rho_{D}] + \frac{ \Gamma a_{1}^{2} }{ 2 } \Lambda ( \sigma_{mm} ) \rho_{D} \nonumber \\
  &\quad  + \frac{ \Gamma a_{5}^{2} }{ 2 } \Lambda ( \sigma_{uu} ) \rho_{D} + \frac{ \Gamma a_{9}^{2} }{ 2 } \Lambda ( \sigma_{ll} ) \rho_{D} \nonumber \\
  &\quad + \frac{ \Gamma a_{2}^{2} }{ 2 } \Lambda ( \sigma_{-}^{um} ) \rho_{D} + \frac{ \Gamma a_{4}^{2} }{ 2 } \Lambda ( \sigma_{+}^{um} ) \rho_{D} \nonumber \\
  &\quad + \frac{ \Gamma a_{7}^{2} }{ 2 } \Lambda ( \sigma_{-}^{ml} ) \rho_{D} + \frac{ \Gamma a_{3}^{2} }{ 2 } \Lambda ( \sigma_{+}^{ml} ) \rho_{D} \nonumber \\
  &\quad + \frac{ \Gamma a_{8}^{2} }{ 2 } \Lambda ( \sigma_{-}^{ul} ) \rho_{D} + \frac{ \Gamma a_{6}^{2} }{ 2 } \Lambda ( \sigma_{+}^{ul} ) \rho_{D} \nonumber \\
  &\quad + \Gamma a_{1} a_{5} \left( \sigma_{mm} \rho_{D} \sigma_{uu} + \sigma_{uu} \rho_{D} \sigma_{mm} \right) \nonumber \\
  &\quad + \Gamma a_{1} a_{9} \left( \sigma_{mm} \rho_{D} \sigma_{ll} + \sigma_{ll} \rho_{D} \sigma_{mm} \right) \nonumber \\
  &\quad + \Gamma a_{5} a_{9} \left( \sigma_{uu} \rho_{D} \sigma_{ll} + \sigma_{ll} \rho_{D} \sigma_{uu} \right) ,
\end{align}
with
\begin{subequations}\label{eq:B_3LA_dressed_state_operators}
  \begin{alignat}{3}
    \sigma^{um}_{-} &= \ket{m} \bra{u} , &\quad \sigma^{um}_{+} &= \ket{u} \bra{m} , \\
    \sigma^{ml}_{-} &= \ket{l} \bra{m} , &\quad \sigma^{ml}_{+} &= \ket{m} \bra{l} , \\
    \sigma^{ul}_{-} &= \ket{l} \bra{u} , &\quad \sigma^{ul}_{+} &= \ket{u} \bra{l} , \\
    \sigma_{z}^{D} &= \ket{u} \bra{u} - \ket{l} \bra{l}, &\quad \sigma_{ii} &= \ket{i} \bra{i} .
  \end{alignat}
\end{subequations}
We expand the reduced density operator in the dressed-state basis and find that the off-diagonal components of the reduced density operator are completely uncoupled. We then derive two sets of moment equations for the dressed-state operators $\left( \omega_{ij} = \omega_{i} - \omega_{j} \right)$:
\begin{subequations}\label{eq:B_off_diagonal_moment_equations}
  \begin{align}
    \ddt \langle \sigma_{\pm}^{um} \rangle &= -\left( \frac{ \Gamma_{um} }{ 2 } \mp i \omega_{um} \right) \langle \sigma_{\pm}^{um} \rangle , \\
    \ddt \langle \sigma_{\pm}^{ul} \rangle &= -\left( \frac{ \Gamma_{ul} }{ 2 } \mp i \omega_{ul} \right) \langle \sigma_{\pm}^{ul} \rangle , \\
    \ddt \langle \sigma_{\pm}^{ml} \rangle &= -\left( \frac{ \Gamma_{ml} }{ 2 } \mp i \omega_{ml} \right) \langle \sigma_{\pm}^{ml} \rangle ,
  \end{align}
\end{subequations}
where
\begin{subequations}
  \begin{align}
    \Gamma_{um} &\equiv \Gamma \left[ a_{2}^{2} + a_{4}^{2} + a_{7}^{2} + a_{8}^{2} + \left( a_{1} - a_{5} \right)^{2} \right] , \\
    \Gamma_{ml} &\equiv \Gamma \left[ a_{3}^{2} + a_{4}^{2} + a_{6}^{2} + a_{7}^{2} + \left( a_{1} - a_{9} \right)^{2} \right] , \\
    \Gamma_{ul} &\equiv \Gamma \left[ a_{2}^{2} + a_{3}^{2} + a_{6}^{2} + a_{8}^{2} + \left( a_{5} - a_{9} \right)^{2} \right] .
  \end{align}
\end{subequations}
and
\begin{equation}\label{eq:B_diagonal_moment_equations}
  \ddt
  \begin{pmatrix}
    \langle \sigma_{mm} \rangle \\
    \langle \sigma_{uu} \rangle \\
    \langle \sigma_{ll} \rangle
  \end{pmatrix}
  = \bm{M}
  \begin{pmatrix}
    \langle \sigma_{mm} \rangle \\
    \langle \sigma_{uu} \rangle \\
    \langle \sigma_{ll} \rangle
  \end{pmatrix} ,
\end{equation}
with
\begin{equation}\label{eq:B_diagonal_moment_equation}
  \bm{M} =
  \begin{pmatrix}
    -\Gamma \left( a_{4}^{2} + a_{7}^{2} \right) & \Gamma a_{2}^{2} & \Gamma a_{3}^{2} \\
    \Gamma a_{4}^{2} & -\Gamma \left( a_{2}^{2} + a_{8}^{2} \right) & \Gamma a_{6}^{2} \\
    \Gamma a_{7}^{2} & \Gamma a_{8}^{2} & -\Gamma \left( a_{3}^{2} + a_{6}^{2} \right)
  \end{pmatrix} .
\end{equation}
Equations~(\ref{eq:B_off_diagonal_moment_equations}) have simple solutions due to their uncoupled nature:
\begin{subequations}
  \begin{align}
    \langle \sigma_{\pm}^{um}(t) \rangle &= \langle \sigma_{\pm}^{um}(0) \rangle e^{ -\left( \frac{ \Gamma_{um} }{ 2 } \mp i \omega_{um} \right) t } , \\
    \langle \sigma_{\pm}^{ul}(t) \rangle &= \langle \sigma_{\pm}^{ul}(0) \rangle e^{ -\left( \frac{ \Gamma_{ul} }{ 2 } \mp i \omega_{ul} \right) t } , \\
    \langle \sigma_{\pm}^{ml}(t) \rangle &= \langle \sigma_{\pm}^{ml}(0) \rangle e^{ -\left( \frac{ \Gamma_{ml} }{ 2 } \mp i \omega_{ml} \right) t } .
  \end{align}
\end{subequations}
Unfortunately, the general solution to Eq.~(\ref{eq:B_diagonal_moment_equations}) is complicated. Taking the limit $\Omega \rightarrow \infty$, however, allows us derive a set of relatively simple solutions:
\begin{subequations}
  \begin{align}
    \langle \sigma_{mm}(t) \rangle &= C_{1} - 2 C_{2} e^{\lambda_{+} t}  , \\
    \langle \sigma_{uu}(t) \rangle &= C_{1} + C_{2} e^{\lambda_{-} t} - C_{3} e^{\lambda_{+} t} , \\
    \langle \sigma_{ll}(t) \rangle &= C_{1} + C_{2} e^{\lambda_{-} t} + C_{3} e^{\lambda_{+} t} ,
  \end{align}
\end{subequations}
where
\begin{subequations}
  \begin{align}
    C_{1} &= \frac{ 1 }{ 3 } \Big( \langle \sigma_{ll}(0) \rangle + \langle \sigma_{uu}(0) \rangle + \langle \sigma_{mm}(0) \rangle \Big) , \\
    C_{2} &= \frac{ 1 }{ 6 } \Big( \langle \sigma_{ll}(0) \rangle + \langle \sigma_{uu}(0) \rangle - 2 \langle \sigma_{mm}(0) \rangle \Big) , \\
    C_{3} &= \frac{ 1 }{ 2 } \Big( \langle \sigma_{ll}(0) \rangle - \langle \sigma_{uu}(0) \rangle \Big) ,
  \end{align}
\end{subequations}
and
\begin{subequations}
  \begin{align}
    \lambda_{-} &\equiv \frac{ -3 \xi^{2} \Gamma }{ 2 \left( 1 + \xi^{2} \right) } , \\
    \lambda_{+} &\equiv \frac{ -\Gamma \left( 1 - \xi^{2} + \xi^{4} \right) }{ 2 \left( 1 + \xi^{2} \right) } .
  \end{align}
\end{subequations}

\bibliography{./Bibliography}

\end{document}